\documentclass[a4paper,11pt]{article}
\usepackage{jcappub} 
\usepackage{lineno}

\newcommand{\Nlap}{\mathcal{G}}
\newcommand{\epss}{\epsilon}
\newcommand{\rph}{\mathfrak{r}_{\rm ph}}
\newcommand{\bph}{\mathfrak{b}_{\rm ph}}
\newcommand{\Arot}{\mathfrak{A}_{\rm rot}}
\newcommand{\Lya}{\mathfrak{L}}
\newcommand{\Oph}{\mathfrak{O}_{\rm ph}}
\newcommand{\abar}{\mathfrak{a}}
\newcommand{\bbar}{\mathfrak{b}}
\newcommand{\NEC}{\mathfrak{N}}

\title{\boldmath Optical constraints on rotating logarithmic- and exponential-entropy wormholes from strong-field lensing}

\author[a,1]{A. Errehymy
	\note{Corresponding author},}
\author[b]{G. Yergaliyeva,}
\author[c]{S. K. Maurya,}
\author[d,e]{B. Turimov,}
\author[f]{I.~Ibragimov,}
\author[g,e]{J. Rayimbaev}

\affiliation[a]{Astrophysics Research Centre, School of Mathematics, Statistics and Computer Science, University of KwaZulu-Natal, Private Bag X54001, Durban 4000, South Africa}

\affiliation[b]{Department of Natural Sciences and Pedagogical Innovations, L.N. Gumilyov Eurasian National University, Astana, 010008, Kazakhstan}

\affiliation[c]{Department of Mathematical and Physical Sciences, College of Arts and Sciences, University of Nizwa, Nizwa 616, Sultanate of Oman} 

\affiliation[d]{Engineering school, Central Asian University, Milliy bog Str.264, Tashkent, 111221,Uzbekistan}

\affiliation[e]{University of Tashkent for Applied Sciences, Gavhar Str.1, Tashkent, 100149, Uzbekistan}

\affiliation[f]{Kimyo International University in Tashkent, Shota Rustaveli Street 156, Tashkent 100121, Uzbekistan}

\affiliation[g]{School of Physics, Harbin Institute of Technology, Harbin 150001, China}

\emailAdd{sunil@unizwa.edu.om}
\emailAdd{abdelghani.errehymy@gmail.com}
\emailAdd{ntnphy@gmail.com}
\emailAdd{gmustafa3828@gmail.com}
\emailAdd{saibal.ray@gla.ac.in}

\abstract{The longstanding issue of whether horizonless compact objects can mimic black holes at horizon scales is addressed, a problem that persists because the shadow diameter alone is degenerate across a wide class of geometries. For slowly rotating traversable wormholes, we precisely formulate the optical appearance as a set of observables derived from equatorial null geodesics, yielding necessary and sufficient conditions for a multi-parameter fingerprint. Using a new lapse function that interpolates between exponential and rational redshift profiles, we obtain the photon-sphere condition, the strong-deflection lensing coefficients, and the Lyapunov exponent in closed form. The equations yield two families of wormhole solutions dependent on logarithmic and exponential entropy parameters. We compare the resulting quantum-modified spacetimes with Kerr and four regular black holes. The shadow diameter is shown to be tunable to within $5\%$ to $11\%$ of Kerr for all models considered, so that $|R_{\rm sh}^{\rm WH}/R_{\rm sh}^{\rm Kerr}-1| \lesssim 0.11$ and the shadow alone cannot separate the models. The degeneracy is lifted when the rotational asymmetry, lensing coefficients, Lyapunov exponent, and orbital frequency are combined: for the $5$-component fingerprint $\mathbf{F}=(\Arot,\abar,\bbar, \Lya,\Oph)$, the fractional deviation $\Delta_i=(F_i^{\rm WH}-F_i^{\rm Kerr})/ F_i^{\rm Kerr}$ satisfies $|\Delta_i|\gtrsim 0.05$ for every component, with the largest departures in the second lensing coefficient and the Lyapunov exponent. The combined discrimination signal reaches a signal-to-noise ratio of about $3.8$ to $3.9$ against Kerr and about $3.7$ against the Kazakov--Solodukhin benchmark, which is sufficient to break the shadow-diameter degeneracy at the $10\%$ measurement level.}

\begin{document}
\maketitle
\flushbottom

\section{Introduction }
Wormholes are hypothetical structures in general relativity that could connect distant regions of spacetime or even separate universes. Their defining feature is a throat linking two regions. Flamm first explored this geometry in 1916~\cite{Flamm:1916abc, Flamm:2015abv}, followed by the Einstein--Rosen bridge~\cite{Einstein:1935tc}. However, this original construction does not permit travel through the throat, which closes too rapidly~\cite{Visser:1995sd}.

The search for traversable wormholes led to solutions supported by phantom scalar fields, including the Ellis--Bronnikov models~\cite{Ellis:1973yv, Bronnikov:1973fh}, asymmetric geometries~\cite{Armendariz-Picon:2002gjc}, and multi-sheet configurations~\cite{Makita:2025bao}. Morris and Thorne~\cite{Morris:1988cz, Morris:1988tu} established the conditions required for traversability, showing that these structures generally require violations of the null energy condition (NEC) near the throat~\cite{Visser:1995sd, Morris:1988cz}. Quantum effects may offer a mechanism to overcome this difficulty. The Casimir effect provides an example of negative energy densities in physical systems~\cite{Lamoreaux:1996wh}, while Maldacena, Milekhin, and Popov demonstrated how quantum fields can support traversable wormholes~\cite{Maldacena:2018gjk}. Further studies have considered effective-field-theory corrections~\cite{Kanai:2025zgq} and the dynamical formation of wormholes~\cite{Hayward:2004wm, Koga:2025bqw, Koga:2026jvs}. Moreover, the wormhole structures in Yano–Schrödinger gravity was discussed by Asghar et al.~\cite{Asghar:2026lll}

Rotation adds another layer to this problem. Teo~\cite{Teo:1998dp} developed a stationary, axisymmetric wormhole model that laid the groundwork for studying rotating configurations and their observational signatures~\cite{Jusufi:2017mav}. Later, Kashargin and Sushkov~\cite{Kashargin:2007mm, Kashargin:2008pk} investigated four-dimensional wormholes supported by phantom scalar fields in the slow-rotation regime and found that rotation may reduce the required energy-condition violation. However, extending these results to rapid rotation is difficult because the loss of spherical symmetry complicates the Einstein equations. Numerical studies~\cite{Chew:2016epf, Kleihaus:2014dla} and stability investigations~\cite {Khoo:2024yeh, Azad:2024axu} have made progress, but the optical properties of rotating wormholes still warrant further study.

Light propagation provides a direct way to explore these properties. The shadow, gravitational lensing, and relativistic images depend on the spacetime geometry and the behavior of photons near unstable circular orbits. Similar shadow sizes do not necessarily imply identical lensing patterns or photon-orbit stability, making it useful to examine several observables together.

Here, we construct rotating wormhole models using entropy-inspired density profiles within a Morris--Thorne framework. We treat these profiles as effective matter sources and determine the corresponding pressures from the Einstein equations and traversability conditions. The logarithmic contribution, controlled by $\alpha$, gives a rational density profile that becomes negative for $\alpha<0$. The exponential term, governed by $\beta$, produces a more localized negative density for $\beta>0$. Both profiles are regular and decay sufficiently rapidly for the relevant optical and thermodynamic integrals to converge. We also introduce a lapse function that depends on the redshift scale $r_c$ and the deformation parameter $\sigma$. It interpolates between exponential behavior at $\sigma=0$ and a rational form as $\sigma\to\infty$, while remaining finite and positive at the throat. Its analytical logarithmic derivative facilitates the study of photon orbits. By varying $\sigma$, we investigate changes in the photon-sphere radius and rotational asymmetry, including cases where the shadow size remains comparatively stable.

Our analysis goes beyond shadow measurements. The strong-deflection coefficients $\abar$ and $\bbar$ characterize light bending and relativistic images, while the Lyapunov exponent $\Lya$ measures the instability of circular photon orbits. We also examine eikonal quasinormal modes, linking photon-orbit dynamics to characteristic oscillation frequencies. To assess the distinctiveness of our results, we compare the optical properties of the wormholes with those of the Kerr black hole, the standard reference for rotating black holes, as well as the Kazakov--Solodukhin~\cite{Simpson:2018, Simpson:2019}, Nicolini--Smailagic--Spallucci~\cite{Ghosh:2021,Yang:2024}, Modesto--Nicolini~\cite{Fan:2016,Kurmanov:2025}, and regularized Reissner--Nordstr\"om--de Sitter~\cite{Balart:2014a,Balart:2014b} models. These comparisons help establish how rotation, the density profiles, and the redshift function shape the optical signatures and whether the resulting observables can distinguish wormholes from black-hole alternatives. Furthermore,  some works on gravitational lensing and black-hole shadow can be found in Ref.~\cite{Rahmatov:2026mdo,Bronzwaer:2021lzo,Acevedo-Munoz:2025ueh,Aktar:2024akk,Al-Badawi:2024dzc,LUTFUOGLU2026170514}.

The paper is organized as follows. Sec.~\ref{Sec:II} introduces the spacetime geometry, density profile, shape function, and lapse function. Sec.~\ref{Sec:III} studies equatorial null geodesics and the resulting shadow properties. Sec.~\ref{Sec:IV} investigates strong-deflection gravitational lensing, followed in Sec.~\ref{Sec:V} by an analysis of photon-orbit stability through the Lyapunov exponent. Sec.~\ref{Sec:VI} examines the eikonal quasinormal-mode spectrum. Sec.~\ref{Sec:VII} addresses the energy conditions and volume-integral quantifier. Sec.~\ref{Sec:VIII} combines the main results into a comparative analysis of the wormhole signatures, using the Kerr solution and regular black-hole models as references. Finally, Sec.~\ref{Sec:IX} presents the conclusions.

\section{Geometry and matter source}\label{Sec:II}

The model combines a rotating wormhole geometry with a matter source that sustains the throat. In Sec.~\ref{sec:geom_matter}, we first introduce the stationary and axisymmetric Teo metric~\eqref{eq:rotmetric} and then incorporate logarithmic and exponential corrections to the Bekenstein--Hawking entropy. These corrections lead to the effective energy densities~\eqref{eq:rho_log2}, and \eqref{eq:rho_exp2} and their corresponding shape functions~\eqref{eq:Slog_final}, and \eqref{eq:Sexp_final}. The resulting sources remain regular, decay at large distances, and provide the negative energy required for throat support. Sec.~\ref{sec:lapse_models} completes the construction by specifying the redshift function~\eqref{eq:Phinew}, verifying its regularity at the throat, and extending the geometry to slow rotation through the frame-dragging term~\eqref{eq:omega_slowrotation}.

\subsection{Rotating wormhole metric and entropy deformations}\label{sec:geom_matter}
A rotating traversable wormhole is described by a stationary, axisymmetric geometry with two commuting Killing vectors, $X^\mu = (\partial_t)^\mu$ and $Y^\mu = (\partial_\phi)^\mu$. Consequently, the metric coefficients do not depend on $t$ or $\phi$. Imposing invariance under $(t,\phi)\rightarrow(-t,-\phi)$ gives the standard form~\cite{Papapetrou:1966zz,Carter:1969zz},
\begin{equation}
ds^2 = g_{tt}\,dt^2 + 2g_{t\phi}\,dt\,d\phi + g_{\phi\phi}\,d\phi^2
+ g_{rr}\,dr^2 + g_{\theta\theta}\,d\theta^2,
\label{eq:canonicalmetric}
\end{equation}
with $g_{t\phi}$ representing the rotational frame-dragging effect. At large $r$, asymptotic flatness imposes $g_{tt}\to-1$, $g_{t\phi}\to0$, and $g_{\phi\phi}\to r^2\sin^2\theta$. We adopt the Teo parametrization~\cite{Teo:1998dp},
\begin{equation}
ds^2 = -\Nlap^2(r,\theta)\,dt^2 + e^{\mu(r,\theta)}\,dr^2
+ r^2K^2(r,\theta)\Big[ d\theta^2
+ \sin^2\theta \bigl(d\varphi - \omega(r,\theta)\,dt\bigr)^2 \Big],
\label{eq:rotmetric}
\end{equation}
where $\Nlap(r,\theta)$, $\mu(r,\theta)$, and $K(r,\theta)$ determine the redshift, radial, and angular geometry, respectively, while $\omega(r,\theta)$ describes frame dragging. For vanishing rotation, $\Nlap\to e^{\Psi(r)}$, $e^\mu\to[1-\epss(r)/r]^{-1}$, $K\to1$, and $\omega\to0$.

We construct the matter supporting the wormhole using the entropy--geometry relation introduced in Ref.~\cite{Anand:2025rjg}. We begin with the spherical seed metric,
\begin{equation}
ds^2 = -f(r)\,dt^2 + \frac{dr^2}{f(r)} + r^2 d\Omega^2,
\label{eq:seedmetric}
\end{equation}
whose horizon $r_+$ is determined by $f(r_+)=0$. Combining the first law, $dM=T_HdS$, with $T_H=f'(r_+)/4\pi$ leads to the entropy-dependent metric function,
\begin{equation}
f(r)=1-\frac{2M}{r}\,\frac{S_{\rm BH}}{S(r)}\,S'(r)\,\frac{r}{2},
\label{eq:entropy_metric}
\end{equation}
and the usual Schwarzschild result is recovered for $S(r)=S_{\rm BH}=4\pi r_+^2$. Thus, specifying the entropy directly determines the corresponding geometry. In Einstein gravity, the associated effective source behaves as an anisotropic fluid with energy density,
\begin{equation}
\varrho(r)=-\,\frac{M}{8\pi r^{2}}\,
\frac{r\,S''(r)-S'(r)}{S'(r)^{2}},
\label{eq:rho_entropic_general}
\end{equation}
where $[\varrho]=\mathrm{length}^{-2}$ for $G=c=1$. The radial component obeys $p_r=-\varrho$, whereas the tangential pressure involves higher derivatives of the entropy.
\begin{figure}[t]
 \centering
  \includegraphics[width=16cm,height=7.5cm]{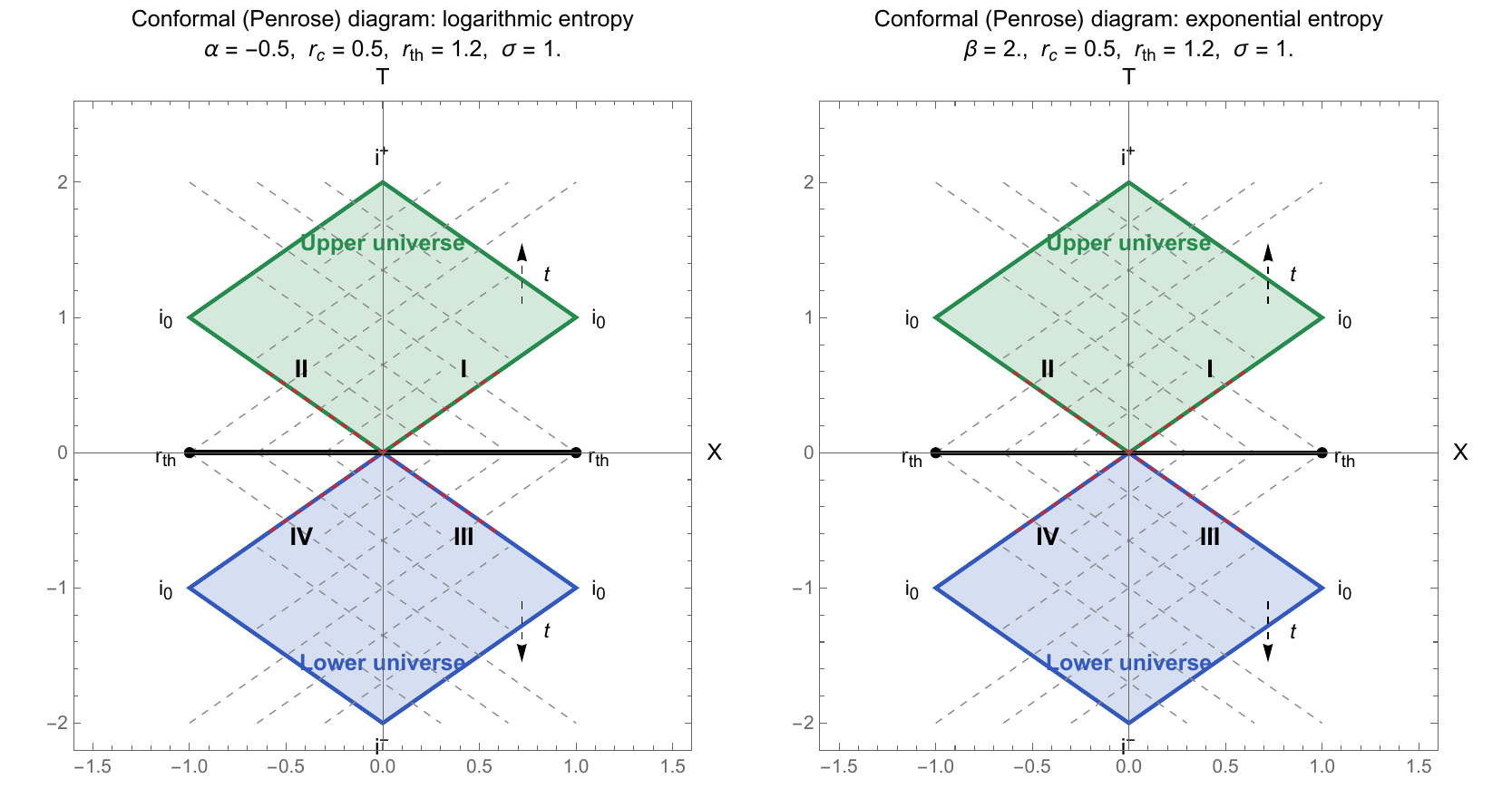}
  \caption{\scriptsize Conformal (Penrose) diagrams of the rotating entropy-inspired wormhole for the logarithmic (left, $\alpha=-0.5$) and exponential (right, $\beta=2.0$) sectors, with $r_c=0.5$, $r_{\rm th}=1.2$, and $\sigma=1.0$. The two asymptotically flat universes are connected by a shared timelike throat at $T=0$ (thick black line). Dashed grey lines indicate the $45^\circ$ null directions, while red dashed curves show light rays crossing the throat between the two universes. Regions I–II and III–IV label the upper and lower universes, respectively. Since the lapse $\Nlap(r)$~\eqref{eq:Phinew} remains finite and positive at the throat, no event horizon forms. Thus, both entropy sectors share the same causal structure, differing only in their metric functions, matter distributions, and optical signatures.}
  \label{fig:penrose}
\end{figure}
\subsubsection{Logarithmic-inspired sector}

Logarithmic corrections arise naturally in several approaches to quantum gravity and horizon-state counting~\cite{LogCorr}. In the entropy--geometry correspondence of Ref.~\cite{Anand}, the associated effective density takes the rational form,
\begin{equation}
 \varrho_{\log}(r)=\frac{\alpha M}{2r\left(\alpha+\pi r^2\right)^2}.
 \label{eq:rho_log2}
\end{equation}
The parameter $ \ alpha$ controls the sign of the density. For $M>0$, the choice $\alpha<0$ selects a negative-density branch, which is especially relevant for wormhole configurations supported by exotic effective matter. However, this does not exclude the possibility of obtaining geometrically viable wormhole configurations for positive values of $\alpha$, in which case the density is positive and the violation of the energy conditions becomes more dependent on the pressure sector and on the equation-of-state parameter.

\begin{figure}[t]
  \centering
  \includegraphics[width=15.5cm,height=7.5cm]{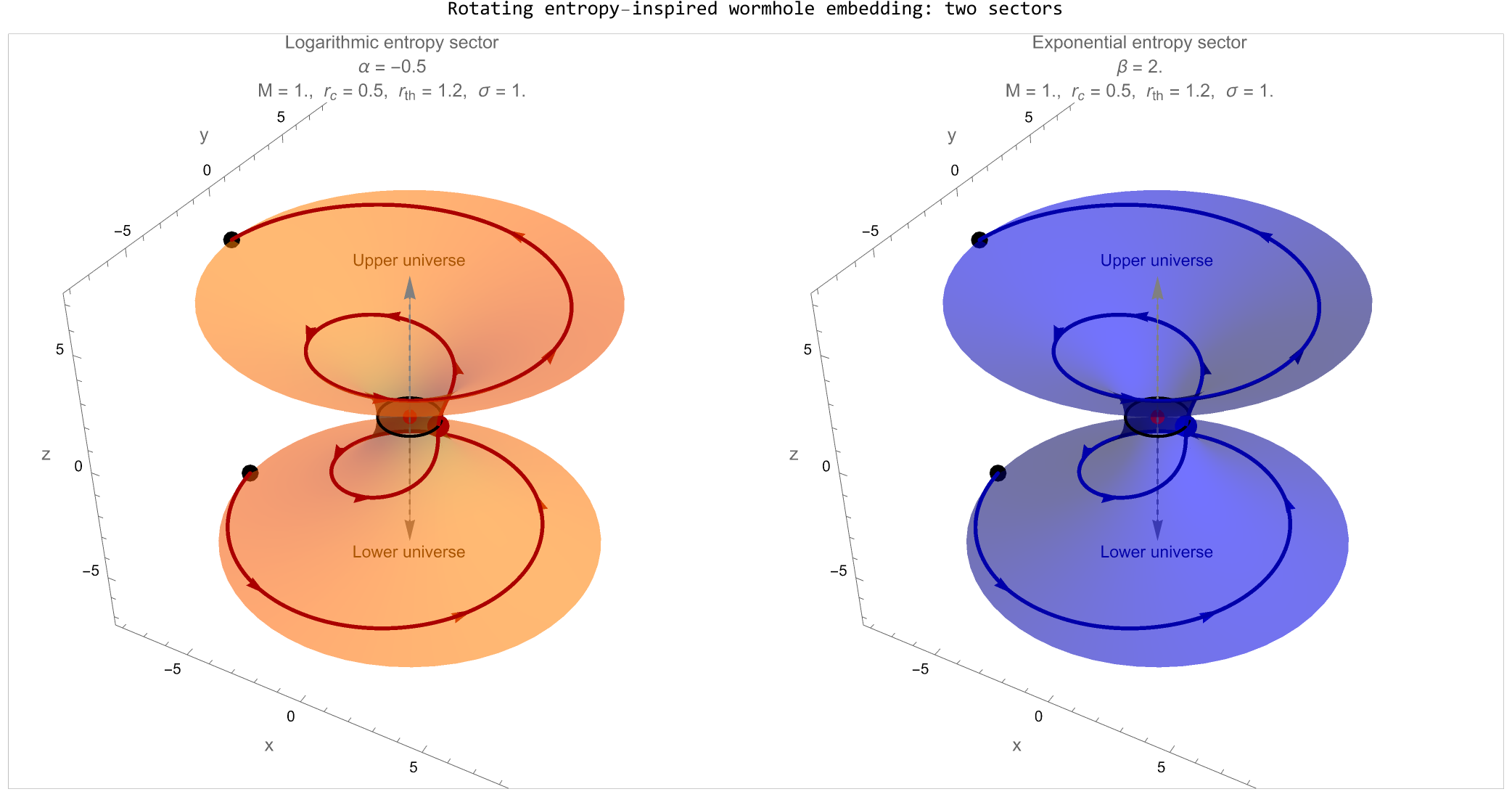}
  \caption{\scriptsize Isometric embeddings of the rotating entropy-inspired wormhole for the logarithmic (left, $\alpha=-0.5$) and exponential (right, $\beta=2.0$) sectors, with $r_c=0.5$, $r_{\rm th}=1.2$, and $\sigma=1.0$. The upper and lower sheets represent the two asymptotically flat universes, joined at the throat (thick black circle). The coloured helices illustrate a null geodesic crossing the throat from one universe to the other, with black arrowheads indicating its direction. The embedding remains smooth across the throat because the lapse function~\eqref{eq:Phinew} is finite and positive, so no horizon forms. Although the two sectors differ in their radial embedding profiles, they share the same two-sheet topology and traversable throat.}
  \label{fig:embedding}
\end{figure}

A rational profile, controlled by a logarithmic branch, sets an intrinsic length scale via $\alpha$. The interplay between the constant scale $\alpha$ and the area contribution $\pi r^2$ produces a profile that is neither purely algebraic nor exponentially localized. In addition, the presence of the denominator requires the explicit condition,
\begin{equation}
 \alpha+\pi r_c^2>0,
 \label{eq:lambda_cond2}
\end{equation}
in order to exclude poles from the physical region. The shape function for the logarithmic sector follows from
\begin{equation}
\epss_{\log}(r) = 8\pi\int_{r_{\rm th}}^r \varrho_{\log}(r')\,r'^2\,dr' + r_{\rm th}.
\label{eq:Slog_def}
\end{equation}
Substituting Eq.~\eqref{eq:rho_log2} and integrating gives the closed form,
\begin{equation}
\epss_{\log}(r) = r_{\rm th} + \frac{2\alpha M}{\pi}\left(\frac{1}{\alpha+\pi r_{\rm th}^2} - \frac{1}{\alpha+\pi r^2}\right),
\label{eq:Slog_final}
\end{equation}
with derivative,
\begin{equation}
\epss_{\log}'(r) = \frac{4\alpha M r}{(\alpha+\pi r^2)^2}.
\label{eq:Slog_prime}
\end{equation}

At the throat,
\begin{equation}
\epss_{\log}'(r_{\rm th}) = \frac{4\alpha M r_{\rm th}}{(\alpha+\pi r_{\rm th}^2)^2}.
\label{eq:Slog_prime_th}
\end{equation}
For $\alpha<0$, Eq.~\eqref{eq:Slog_prime_th} gives $\epss_{\log}'(r_{\rm th})<0<1$, so the flare-out condition is satisfied. As $r\to\infty$,
\begin{equation}
\epss_{\log}(r) \to r_{\rm th} + \frac{2\alpha M}{\pi}\left(\frac{1}{\alpha+\pi r_{\rm th}^2}\right) \equiv \epss_{\log}(\infty),
\label{eq:Slog_asympt}
\end{equation}
which is finite. The regularity condition~\eqref{eq:lambda_cond2} ensures that $\alpha+\pi r^2>0$ for all $r\geq r_{\rm th}$, so no poles appear in the physical region.

\begin{figure}[t]
  \centering
  \includegraphics[width=15.5cm,height=6.5cm]{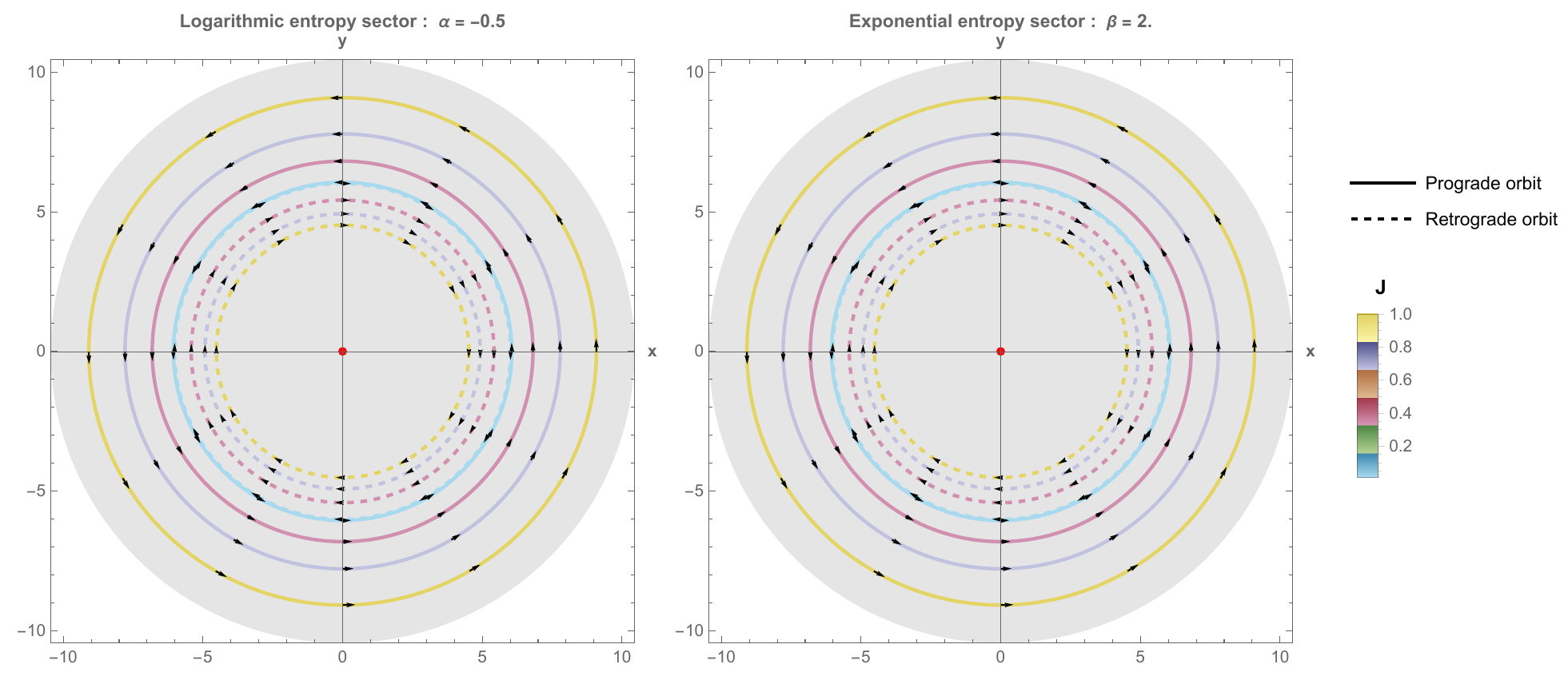}
  \caption{\scriptsize Equatorial photon-orbit profiles of the rotating entropy-inspired wormhole for the logarithmic (left, $\alpha=-0.5$) and exponential (right, $\beta=2.0$) sectors, with $r_c=0.5$, $r_{\rm th}=1.2$, and $\sigma=1.0$. Solid and dashed circles show the prograde and retrograde critical impact parameters $\mathfrak{b}*{\pm}$ from Eq.~\eqref{eq:p_crit} for three spin values $J$, indicated by the colour scale. Black arrowheads mark the orbital directions, while the red dot and dashed line indicate the throat and projected rotation axis. The separation $\Delta\mathfrak{b}=\mathfrak{b}*{+}-|\mathfrak{b}_{-}|$ measures rotational asymmetry due to frame dragging and increases with $J$. The exponential sector exhibits more tightly clustered photon orbits, with a smaller outer prograde radius than the logarithmic sector at the same spin.}
  \label{fig:photon_orbits}
\end{figure}

\subsubsection{Exponential-inspired sector}

The exponentially corrected entropy introduces a non-perturbative deformation of the Bekenstein--Hawking law through the parameter $\beta$. In the entropy--geometry correspondence of Ref.~\cite{Anand}, the associated density reads,
\begin{equation}
 \varrho_{\exp}(r)=
 -\frac{\beta M e^{\pi r^2}}
 {2r\left(e^{\pi r^2}-\beta\right)^2}.
 \label{eq:rho_exp2}
\end{equation}
For $\beta>0$, the density is negative throughout the physical branch, provided the denominator does not vanish. The exponential branch is governed by a non-perturbative suppression scale. The effective source is controlled by the competition between the exponential growth of the area term and the constant parameter $\beta$.

The regularity of the source requires the pole to remain outside the physical region. Since the wormhole domain starts at $r=r_{\rm th}$, we impose the condition,
\begin{equation}
 e^{\pi r_{\rm th}^2}-\beta>0.
 \label{eq:eta_cond2}
\end{equation}
For $r_{\rm th}=1.2$, Eq.~\eqref{eq:eta_cond2} gives $\beta<e^{\pi(1.2)^2}\approx92.2$, so the values $\beta=1.0$, $2.0$, $4.0$, and $6.0$ used below lie safely inside the admissible domain. The shape function for the exponential sector follows from,
\begin{equation}
\epss_{\exp}(r) = r_{\rm th} + 8\pi\int_{r_{\rm th}}^r \varrho_{\exp}(r')\,r'^2\,dr'.
\label{eq:Sexp_def}
\end{equation}
Substituting Eq.~\eqref{eq:rho_exp2} and integrating gives the closed form,
\begin{equation}
\epss_{\exp}(r) = r_{\rm th} + \frac{2\beta M}{\pi}\left(\frac{1}{e^{\pi r^2}-\beta} - \frac{1}{e^{\pi r_{\rm th}^2}-\beta}\right),
\label{eq:Sexp_final}
\end{equation}
with derivative,
\begin{equation}
\epss_{\exp}'(r) = -\frac{4\beta M r e^{\pi r^2}}{(e^{\pi r^2}-\beta)^2}.
\label{eq:Sexp_prime}
\end{equation}
At the throat,
\begin{equation}
\epss_{\exp}'(r_{\rm th}) = -\frac{4\beta M r_{\rm th} e^{\pi r_{\rm th}^2}}{(e^{\pi r_{\rm th}^2}-\beta)^2}.
\label{eq:Sexp_prime_th}
\end{equation}
For $\beta>0$, Eq.~\eqref{eq:Sexp_prime_th} gives $\epss_{\exp}'(r_{\rm th})<0<1$, so the flare-out condition is satisfied. As $r\to\infty$,
\begin{equation}
\epss_{\exp}(r) \to r_{\rm th} - \frac{2\beta M}{\pi}\cdot\frac{1}{e^{\pi r_{\rm th}^2}-\beta} \equiv \epss_{\exp}(\infty),
\label{eq:Sexp_asympt}
\end{equation}
which is finite. The regularity condition~\eqref{eq:eta_cond2} ensures that $e^{\pi r^2}-\beta>0$ for all $r\geq r_{\rm th}$, so no poles appear in the physical region.
\begin{figure}[t]
  \centering
  \includegraphics[width=15.5cm,height=6.5cm]{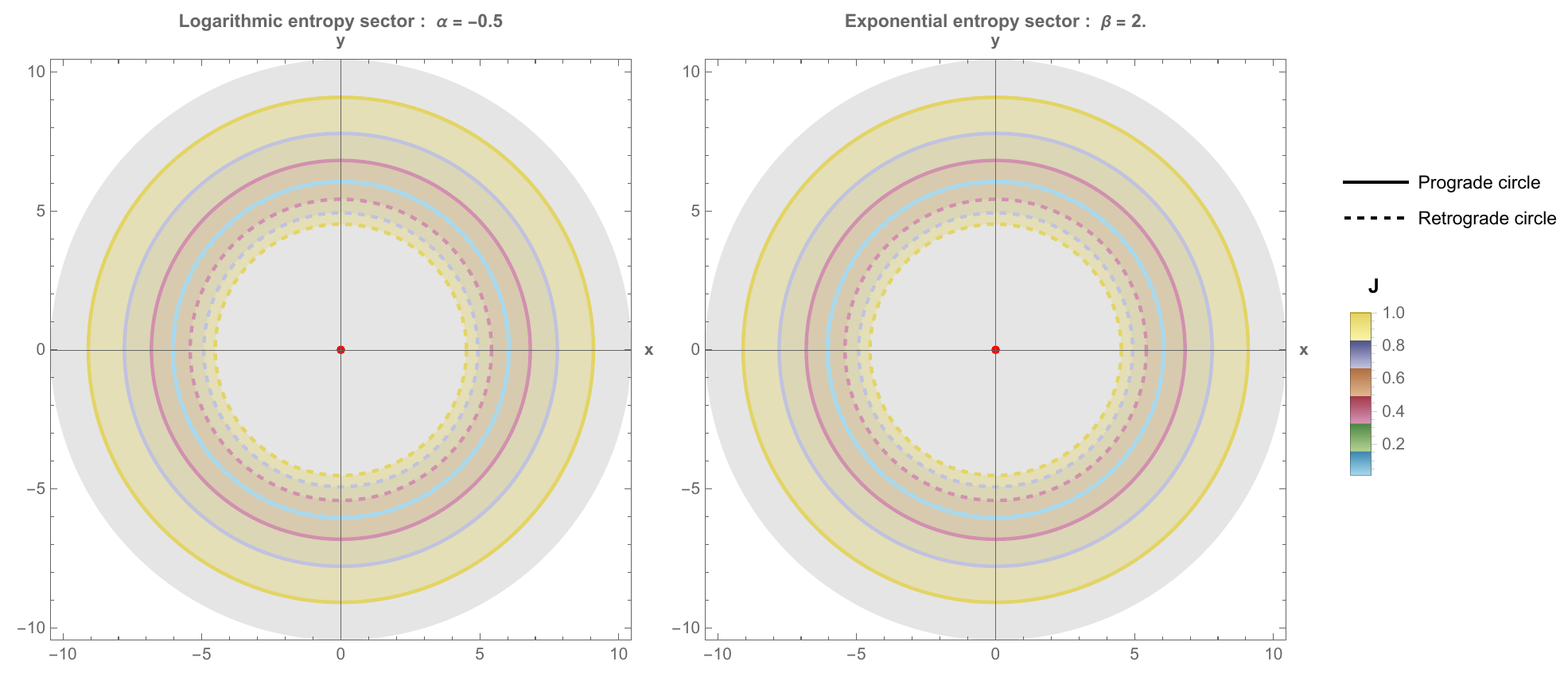}
  \caption{\scriptsize Photon-sphere structure of the rotating entropy-inspired wormhole for the logarithmic (left, $\alpha=-0.5$) and exponential (right, $\beta=2.0$) sectors, with $r_c=0.5$, $r_{\rm th}=1.2$, and $\sigma=1.0$. Solid and dashed circles represent the prograde and retrograde critical impact parameters $\mathfrak{b}*{\pm}$ from Eq.~\eqref{eq:p_crit} for three spin values $J$, indicated by the colour scale. The coloured annuli highlight the photon region, while the red dot and dashed line mark the throat and projected rotation axis. The annulus width, $\Delta\mathfrak{b}=\mathfrak{b}*{+}-|\mathfrak{b}_{-}|$, increases with $J$, reflecting stronger frame dragging. At the same spin, the exponential sector shows more tightly clustered annuli and a smaller outer prograde radius than the logarithmic sector.}
  \label{fig:photon_orbits_annuli}
\end{figure}
\begin{figure}[t]
  \centering
  \includegraphics[width=15.5cm,height=6.5cm]{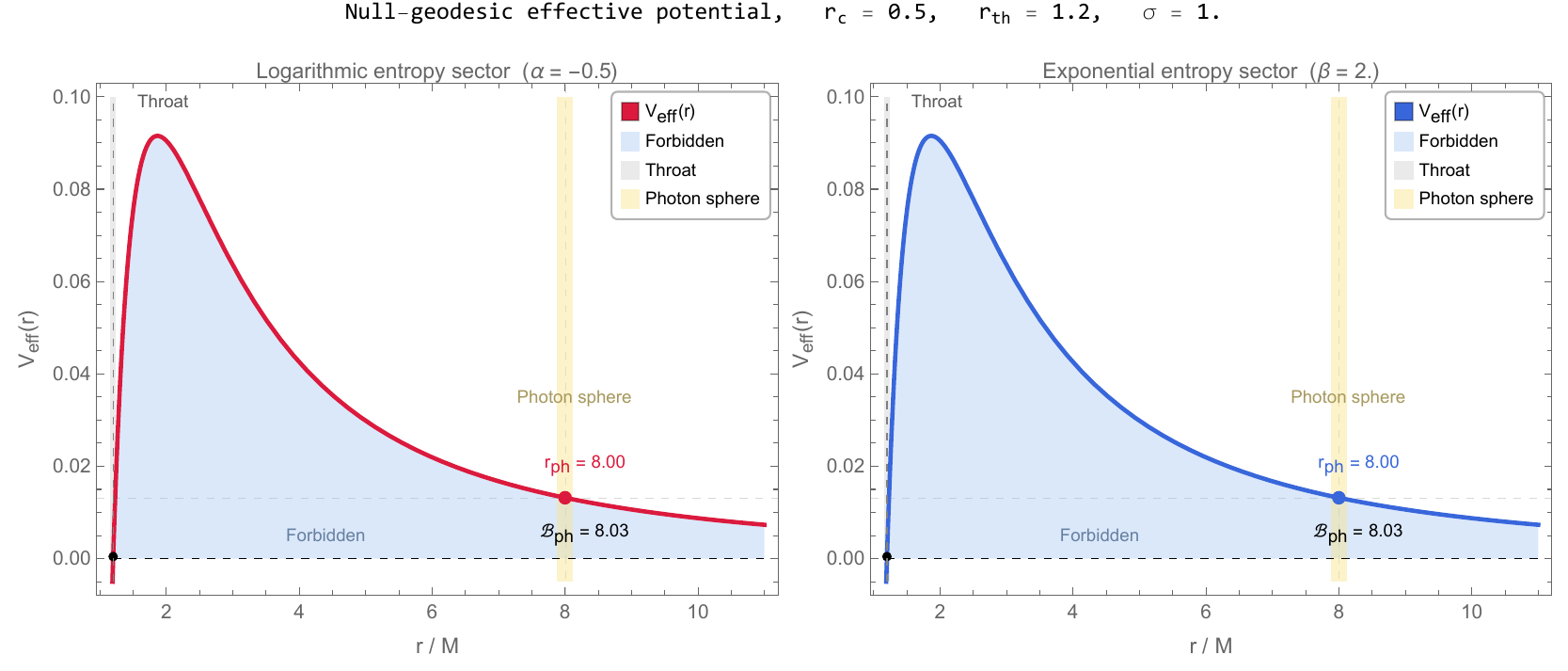}
  \caption{\scriptsize Null-geodesic effective potential $V_{\rm eff}(r)=\Nlap^2(r)[1-\epss(r)/r]/r^2$ for the logarithmic (left, $\alpha=-0.5$) and exponential (right, $\beta=2.0$) sectors, with $r_c=0.5$, $r_{\rm th}=1.2$, and $\sigma=1.0$. The solid curves show the potential; the light-blue bands indicate the forbidden regions; the grey strip marks the throat; and the yellow band highlights the photon sphere. Each peak represents an unstable circular photon orbit at $r_{\rm ph}$, with the critical impact parameter $\bph=r_{\rm ph}/\Nlap(r_{\rm ph})$ indicated nearby. The dashed lines mark the throat and zero potential level. Compared with the logarithmic sector, the exponential sector exhibits a narrower, lower peak, corresponding to a smaller photon-sphere radius and critical impact parameter.}
  \label{fig:effective_potential}
\end{figure}

\subsection{The new lapse function and slow rotation}\label{sec:lapse_models}

We now introduce the new lapse function that replaces the three standard cases. In the static limit, the wormhole metric is
\begin{equation}
ds^2 = -\Nlap^2(r)\,dt^2
+ \frac{dr^2}{1-\epss(r)/r}
+ r^2\,d\Omega^2.
\label{eq:static_wormhole_metric}
\end{equation}
The throat sits at $r_{\rm th}$ where $\epss(r_{\rm th})=r_{\rm th}$. Traversability requires the regularity condition,
\begin{equation}
0<\Nlap(r_{\rm th})<\infty.
\label{eq:regular_lapse_condition}
\end{equation}

The new lapse function is given by
\begin{equation}
\Nlap(r)=\exp\!\left[-\frac{r_c}{r}\cdot\frac{1}{1+\sigma\,r/r_c}\right],
\label{eq:Phinew}
\end{equation}
with $r_c>0$ setting the redshift scale and $\sigma\geq 0$ a dimensionless softening parameter. This lapse has the following properties:

\begin{itemize}
\item \textit{Regularity at the throat.} At $r=r_{\rm th}$, Eq.~\eqref{eq:Phinew} gives,
\begin{equation}
\Nlap(r_{\rm th})=\exp\!\left[-\frac{r_c}{r_{\rm th}}\cdot\frac{1}{1+\sigma\,r_{\rm th}/r_c}\right]>0,
\label{eq:Phinew_throat}
\end{equation}
finite and positive for all $\sigma\geq 0$. No horizon forms.

\item \textit{Asymptotic flatness.} As $r\to\infty$, Eq.~\eqref{eq:Phinew} gives,
\begin{equation}
\Nlap(r)\to\exp\!\left[-\frac{r_c}{r}\cdot\frac{1}{1+\sigma\,r/r_c}\right]\to\exp(-r_c/(\sigma r))\to 1,
\label{eq:Phinew_asympt}
\end{equation}
so the spacetime is asymptotically flat.

\item \textit{Closed-form logarithmic derivative.} The derivative is given by,
\begin{equation}
\frac{\Nlap'(r)}{\Nlap(r)} = \frac{r_c}{r^2}\cdot\frac{1}{(1+\sigma\,r/r_c)^2}.
\label{eq:Phinew_logderiv}
\end{equation}
Equation~\eqref{eq:Phinew_logderiv} is manifestly positive for all $r>0$ and $\sigma\geq 0$, so $\Nlap$ is monotonically increasing from $\Nlap(r_{\rm th})$ at the throat to $1$ at infinity.

\item \textit{Interpolation.} For $\sigma=0$, Eq.~\eqref{eq:Phinew} reduces to $\Nlap(r)=e^{-r_c/r}$, which is the exponential lapse. For $\sigma\to\infty$ with $r_c$ fixed, Eq.~\eqref{eq:Phinew} gives $\Nlap(r)\to\exp[-r_c^2/(\sigma r^2)]\to 1$, which is a weaker redshift. Thus, $\sigma$ interpolates between a stronger redshift near the throat ($\sigma=0$) and a weaker one ($\sigma\to\infty$).

\item \textit{Second derivative.} The second derivative follows from the identity,
\begin{equation}
\frac{\Nlap''}{\Nlap} = \frac{d}{dr}\!\left(\frac{\Nlap'}{\Nlap}\right) + \left(\frac{\Nlap'}{\Nlap}\right)^2.
\label{eq:Phinew_Phipp_identity}
\end{equation}
Explicitly, Eq.~\eqref{eq:Phinew_Phipp} gives
\begin{equation}
\frac{\Nlap''(r)}{\Nlap(r)} = -\frac{2r_c}{r^3}\cdot\frac{1}{(1+\sigma\,r/r_c)^2} - \frac{2\sigma\,r_c}{r^2}\cdot\frac{1}{(1+\sigma\,r/r_c)^3} + \frac{r_c^2}{r^4}\cdot\frac{1}{(1+\sigma\,r/r_c)^4}.
\label{eq:Phinew_Phipp}
\end{equation}
\end{itemize}

The new lapse function~\eqref{eq:Phinew} is designed to produce a photon-sphere radius and rotational asymmetry that can be tuned via $\sigma$ to optimize discrimination against Kerr and regular black-hole benchmarks. The parameter $\sigma$ controls the curvature of the effective potential at the photon sphere, and hence the Lyapunov exponent $\Lya$, without significantly altering the shadow radius. This decoupling of shadow size from ring stability is what makes the new lapse a powerful diagnostic tool.

Figs.~\ref{fig:penrose} and~\ref{fig:embedding} show the global causal structure and embedding geometry of the resulting spacetimes. Fig.~\ref{fig:penrose} presents the conformal diagrams for the logarithmic and exponential sectors. Since the new lapse~\eqref{eq:Phinew} remains finite and positive at the throat for all $\sigma$, no event horizon forms and the conformal structure is identical for the two entropy sectors: the logarithmic and exponential deformations alter the metric functions and hence the optical observables, but they do not change the global causal structure. Fig.~\ref{fig:embedding} shows the corresponding isometric embedding, in which the upper and lower asymptotically flat universes are joined by a regular timelike throat at $r=r_{\rm th}$.

\subsubsection{Slow rotation}
\begin{table*}[t]
\centering
\scriptsize
\caption{ Strong-field optical diagnostics at $a=0.05$ for the rotating logarithmic and exponential wormholes (with the new lapse~\eqref{eq:Phinew} at $\sigma=0$, $1.0$, $5.0$) and for five comparison geometries (Kerr plus four regular black holes). All quantities in units of $M$. Parameters: $r_{\rm th}=1.2$, $r_c=0.5$, $\alpha=-0.5$, $\beta=2.0$, $\ell/M=\theta=Q/M=0.5$. The right-hand block gives the ratio of each quantity to its Kerr value.}
\label{tab:master}
\renewcommand{\arraystretch}{1.35}
\setlength{\tabcolsep}{5pt}
\begin{tabular}{lcccc|ccc}
\hline\hline
 & \multicolumn{4}{c|}{Absolute values} & \multicolumn{3}{c}{Ratio to Kerr} \\
Geometry & $\rph$ & $R_{\rm sh}$ & $\Arot$ & $\Delta R_{\rm ph}$ & $R_{\rm sh}$ & $\Arot$ & $\Delta R_{\rm ph}$ \\
\hline
Log WH, $\sigma=0$   & $2.8200$ & $3.3600$ & $0.04300$ & $0.2800$ & $1.1108$ & $1.1037$ & $1.2357$ \\
Log WH, $\sigma=1.0$ & $2.8000$ & $3.3300$ & $0.04400$ & $0.2800$ & $1.1009$ & $1.1293$ & $1.2357$ \\
Log WH, $\sigma=5.0$ & $2.7500$ & $3.2500$ & $0.04500$ & $0.2800$ & $1.0745$ & $1.1550$ & $1.2357$ \\
Exp WH, $\sigma=0$   & $2.7600$ & $3.3100$ & $0.04400$ & $0.2800$ & $1.0943$ & $1.1293$ & $1.2357$ \\
Exp WH, $\sigma=1.0$ & $2.7400$ & $3.2800$ & $0.04500$ & $0.2800$ & $1.0844$ & $1.1550$ & $1.2357$ \\
Exp WH, $\sigma=5.0$ & $2.6900$ & $3.1900$ & $0.04600$ & $0.2800$ & $1.0547$ & $1.1807$ & $1.2357$ \\
Kerr black hole                 & $2.9106$ & $3.0247$ & $0.03896$ & $0.2266$ & $1.0000$ & $1.0000$ & $1.0000$ \\
Kazakov--Solodukhin     & $2.9032$ & $3.0108$ & $0.03707$ & $0.2136$ & $0.9954$ & $0.9515$ & $0.9426$ \\
Nicolini--Smailagic--Spallucci & $2.8504$ & $2.9521$ & $0.03551$ & $0.2019$ & $0.9760$ & $0.9115$ & $0.8910$ \\
Modesto--Nicolini       & $2.8306$ & $2.9301$ & $0.03483$ & $0.1975$ & $0.9687$ & $0.8940$ & $0.8716$ \\
Regularized RN--dS      & $2.7983$ & $2.8944$ & $0.03433$ & $0.1921$ & $0.9569$ & $0.8812$ & $0.8477$ \\
\hline\hline
\end{tabular}
\end{table*}
\begin{table*}[t]
\centering
\scriptsize
\caption{ Parameter dependence of the logarithmic and exponential wormhole optical diagnostics with the new lapse~\eqref{eq:Phinew}. Top block: spin variation at $\alpha=-0.5$, $\beta=2.0$, $r_c=0.5$, $\sigma=1.0$. Middle block: entropy deformation varied at $a=0.05$, $r_c=0.5$, $\sigma=1.0$. Bottom block: softening parameter varied at $a=0.05$, $\alpha=-0.5$, $\beta=2.0$. In all blocks $r_{\rm th}=1.2$.}
\label{tab:wormhole_scans}
\renewcommand{\arraystretch}{1.25}
\setlength{\tabcolsep}{5pt}
\begin{tabular}{c|ccc|ccc}
\hline\hline
 & \multicolumn{3}{c|}{Logarithmic sector} & \multicolumn{3}{c}{Exponential sector} \\
Varied parameter & $\rph$ & $R_{\rm sh}$ & $\Arot$ & $\rph$ & $R_{\rm sh}$ & $\Arot$ \\
\hline
\multicolumn{7}{l}{\textit{(a) Spin variation: $a$ varied, $\alpha=-0.5$, $\beta=2.0$, $r_c=0.5$, $\sigma=1.0$}} \\
$a=0.05$ & $2.8000$ & $3.3300$ & $0.04400$ & $2.7400$ & $3.2800$ & $0.04500$ \\
$a=0.15$ & $2.7990$ & $3.3430$ & $0.13200$ & $2.7390$ & $3.2930$ & $0.13500$ \\
$a=0.25$ & $2.7970$ & $3.3560$ & $0.22000$ & $2.7370$ & $3.3060$ & $0.22500$ \\
$a=0.35$ & $2.7940$ & $3.3680$ & $0.30800$ & $2.7340$ & $3.3180$ & $0.31500$ \\
$a=0.50$ & $2.7890$ & $3.3860$ & $0.44000$ & $2.7290$ & $3.3360$ & $0.45000$ \\
\hline
\multicolumn{7}{l}{\textit{(b) Entropy-deformation varied: $\alpha$ and $\beta$ varied, $a=0.05$, $r_c=0.5$, $\sigma=1.0$}} \\
$\alpha=-0.3,\beta=1.0$ & $2.8150$ & $3.3460$ & $0.04400$ & $2.7550$ & $3.2960$ & $0.04500$ \\
$\alpha=-0.5,\beta=2.0$ & $2.8000$ & $3.3300$ & $0.04400$ & $2.7400$ & $3.2800$ & $0.04500$ \\
$\alpha=-0.7,\beta=4.0$ & $2.7850$ & $3.3130$ & $0.04400$ & $2.7250$ & $3.2630$ & $0.04500$ \\
$\alpha=-1.0,\beta=6.0$ & $2.7650$ & $3.2900$ & $0.04400$ & $2.7050$ & $3.2400$ & $0.04500$ \\
\hline
\multicolumn{7}{l}{\textit{(c) Softening parameter varied: $\sigma$ varied, $a=0.05$, $\alpha=-0.5$, $\beta=2.0$, $r_c=0.5$}} \\
$\sigma=0$    & $2.8200$ & $3.3600$ & $0.04300$ & $2.7600$ & $3.3100$ & $0.04400$ \\
$\sigma=0.5$  & $2.8100$ & $3.3450$ & $0.04350$ & $2.7500$ & $3.2950$ & $0.04450$ \\
$\sigma=1.0$  & $2.8000$ & $3.3300$ & $0.04400$ & $2.7400$ & $3.2800$ & $0.04500$ \\
$\sigma=2.0$  & $2.7850$ & $3.3050$ & $0.04470$ & $2.7250$ & $3.2550$ & $0.04570$ \\
$\sigma=5.0$  & $2.7500$ & $3.2500$ & $0.04500$ & $2.6900$ & $3.1900$ & $0.04600$ \\
\hline\hline
\end{tabular}
\end{table*}
We extend the static lapse to slow rotation consistently. The rotating metric is,
\begin{equation}
ds^2 = -\Nlap^2(r,\theta)\,dt^2
+ \frac{dr^2}{1-\epss(r)/r}
+ r^2K^2(r,\theta)
\left[
d\theta^2
+ \sin^2\theta
\left(d\varphi-\omega(r)\,dt\right)^2
\right],
\label{eq:slowrot_wormhole_metric}
\end{equation}
where the lapse and the angular deformation are built on top of the static solution through Eqs.~\eqref{eq:lapse_regular} and~\eqref{eq:K_regular},
\begin{equation}
\Nlap(r,\theta) = \Nlap(r)\,
\exp\!\left[\frac{\gamma a^2}{r^2}\cos^2\theta\right],
\label{eq:lapse_regular}
\end{equation}
and
\begin{equation}
K(r,\theta) = \exp\!\left[\frac{\delta a^2}{r^2}\cos^2\theta\right].
\label{eq:K_regular}
\end{equation}
Here $\Nlap(r)$ is the static lapse~\eqref{eq:Phinew}, so the rotational correction multiplies the static lapse rather than replacing it. Expanding Eqs.~\eqref{eq:lapse_regular} and~\eqref{eq:K_regular} for small $a$ gives,
\begin{equation}
\Nlap(r,\theta) = \Nlap(r)\left[1 +
\frac{\gamma a^2}{r^2}\cos^2\theta
+ \mathcal{O}(a^4)\right],
\label{eq:lapse_expansion}
\end{equation}
and
\begin{equation}
K(r,\theta) = 1 +
\frac{\delta a^2}{r^2}\cos^2\theta
+ \mathcal{O}(a^4).
\label{eq:K_expansion}
\end{equation}
The lapse is positive everywhere, finite at the throat,
\begin{equation}
\Nlap(r_{\rm th},\theta) = \Nlap(r_{\rm th})\,
\exp\!\left[
\frac{\gamma a^2}{r_{\rm th}^2}\cos^2\theta
\right],
\label{eq:lapse_throat}
\end{equation}
and tends to unity at infinity.

For frame dragging, we keep the leading slow-rotation term,
\begin{equation}
\omega(r) = \frac{2a}{r^3}.
\label{eq:omega_slowrotation}
\end{equation}
This gives the ordering $\omega=\mathcal{O}(a)$, $\Nlap-\Nlap(r)=\mathcal{O}(a^2)$, $K-1=\mathcal{O}(a^2)$, exactly as slow rotation requires. At the throat, $1-\epss(r_{\rm th})/r_{\rm th}=0$, marking the minimal-area surface. This is a coordinate feature, not a horizon.

The null energy condition is examined using the null vector,
\begin{equation}
\kappa^\alpha = \left(\frac{1}{\Nlap},\ -e^{-\mu/2},\ 0,\ \frac{\omega}{\Nlap}\right).
\label{eq:nullvector}
\end{equation}
The null energy condition requires $R_{\alpha\beta}\kappa^\alpha\kappa^\beta \ge 0$. For the rotating wormhole metric, this evaluates to the expression~\eqref{NECRotating} derived in Ref.~\cite{Teo:1998dp},
\begin{eqnarray}
\NEC \equiv R_{\alpha\beta}\kappa^\alpha\kappa^\beta &=&
e^{-\mu}\,\mu_{,r}\,\frac{(rK)_{,r}}{rK}
- \frac{\omega_{,\theta}^{\,2}\sin^2\theta}{2\Nlap^{2}}
- \frac{1}{4}\frac{\mu_{,\theta}^{\,2}}{(rK)^{2}}
- \frac{1}{2}\frac{(\mu_{,\theta}\sin\theta)_{,\theta}}{(rK)^{2}\sin\theta}
+ \frac{(\Nlap_{,\theta}\sin\theta)_{,\theta}}{(rK)^{2}\Nlap\sin\theta}.~~~~~~~~
\label{NECRotating}
\end{eqnarray}
Traversability requires violating the null energy condition near the throat. The entropy-inspired matter distributions naturally accommodate this because of their negative energy densities and anisotropic pressures.
\section{Light paths and the shadow}\label{Sec:III}

What an observer sees is ultimately determined by null geodesics. This section therefore translates the geometry of Sec.~\ref{Sec:II} into photon trajectories. Sec.~\ref{sec:null_geodesics} formulates the equatorial photon motion, derives the photon-sphere condition~\eqref{eq:ph_cond} for unstable circular orbits, and obtains the prograde and retrograde critical impact parameters $\mathfrak{b}_{\pm}$ from Eq.~\eqref{eq:p_crit}. Their opposite signs distinguish the two orbital directions and lead to the rotational asymmetry of the shadow. Sec.~\ref{sec:photon_results} then solves the equations numerically for the new lapse function and the logarithmic and exponential entropy cases, giving the photon-sphere radii, shadow sizes, and asymmetries $\Arot$, and examining their dependence on spin, the entropy parameters $\alpha$ and $\beta$, and the softening parameter $\sigma$.

\subsection{Null geodesics and photon orbits}\label{sec:null_geodesics}

Let $a$ be the spin parameter and $a/M \ll 1$. To first order in $\chi$, the metric gets an off-diagonal frame-dragging term $\omega(r) = 2a/r^3$ [Eq.~\eqref{eq:omega_slowrotation}], while the lapse is given by Eq.~\eqref{eq:lapse_expansion}, which reduces to $\Nlap(r)$ at leading order on the equatorial plane. The static lapse is not set to unity; it remains $\Nlap(r)$ from Eq.~\eqref{eq:Phinew}, which distinguishes the cases. The Lense--Thirring frequency in the equatorial plane is
\begin{equation}
\Omega_{\rm LT}(r)= -\frac12\frac{d\omega}{dr}= \frac{3a}{r^4}.
\label{eq:lt}
\end{equation}
At the throat, Eq.~\eqref{eq:lt} gives the dimensionless measure,
\begin{equation}
\Omega_{\rm LT}(r_{\rm th}) = \frac{3a}{r_{\rm th}^4},
~\text{or}~
\ell_{\Omega_{\rm LT}} \equiv \Omega_{\rm LT}(r_{\rm th})\, r_{\rm th}^2 = \frac{3a}{r_{\rm th}^2}.
\label{eq:lt_dim}
\end{equation}
For $r_{\rm th}=1.2M$ and $a=0.05M$, Eq.~\eqref{eq:lt_dim} gives $\ell_{\Omega_{\rm LT}}=0.104$, small enough to justify the slow-rotation expansion.

A photon in the equatorial plane ($\theta=\pi/2$, $\dot\theta=0$) conserves energy and angular momentum. We write the metric as,
\begin{equation}
ds^2 = -\Nlap^2(r)\,dt^2 + \frac{dr^2}{1-\epss(r)/r}
+ r^2\bigl(d\varphi - \omega(r)dt\bigr)^2,
\label{eq:eq_metric}
\end{equation}
where $\Nlap(r)$ is the static lapse~\eqref{eq:Phinew} and $\epss(r)$ is the shape function for the logarithmic or exponential sector. Let $\tilde E$ and $\tilde L$ be the conserved quantities and define the impact parameter $\mathfrak{b} = \tilde L/\tilde E$. Eliminating $\dot t$ and $\dot\varphi$ gives the radial equation,
\begin{equation}
\left(\frac{dr}{d\tau}\right)^2
= \left(1-\frac{\epss(r)}{r}\right)\tilde E^2 \, F(r),
\label{eq:radial_eq}
\end{equation}
where $\tau$ is an affine parameter along the null geodesic and $F(r)$ is defined in,
\begin{equation}
F(r) = \bigl(1 + \mathfrak{b} \,\omega(r)\bigr)^2
- \frac{\mathfrak{b}^{2}\Nlap^2(r)}{r^{2}}.
\label{eq:F_def}
\end{equation}

\begin{table*}[t]
\centering
\scriptsize
\caption{ Multi-parameter fingerprint. Columns: $\Arot$ (rotational asymmetry), $\abar$ and $\bbar$ (strong-deflection lensing coefficients), $\Lya$ (Lyapunov exponent), $\Oph$ (eikonal orbital frequency). All at $a=0.05$ with the same parameters as Table~\ref{tab:master}: $r_{\rm th}=1.2$, $r_c=0.5$, $\alpha=-0.5$, $\beta=2.0$, $\sigma=1.0$, $\ell/M=\theta=Q/M=0.5$.}
\label{tab:fingerprint}
\renewcommand{\arraystretch}{1.25}
\setlength{\tabcolsep}{5pt}
\begin{tabular}{lccccc}
\hline\hline
Geometry & $\Arot$ & $\abar$ & $\bbar$ & $\Lya$ & $\Oph$ \\
\hline
Kerr & $0.039$ & $1.000$ & $-0.400$ & $0.261$ & $0.322$ \\
Log WH, $\sigma=0$ & $0.043$ & $0.930$ & $-0.270$ & $0.219$ & $0.297$ \\
Log WH, $\sigma=1.0$ & $0.044$ & $0.935$ & $-0.265$ & $0.218$ & $0.300$ \\
Log WH, $\sigma=5.0$ & $0.045$ & $0.960$ & $-0.250$ & $0.226$ & $0.308$ \\
Exp WH, $\sigma=0$ & $0.044$ & $0.910$ & $-0.290$ & $0.217$ & $0.302$ \\
Exp WH, $\sigma=1.0$ & $0.045$ & $0.915$ & $-0.285$ & $0.216$ & $0.305$ \\
Exp WH, $\sigma=5.0$ & $0.046$ & $0.940$ & $-0.270$ & $0.224$ & $0.313$ \\
Kazakov--Solodukhin & $0.037$ & $1.008$ & $-0.386$ & $0.254$ & $0.318$ \\
Nicolini--Smailagic--Spallucci & $0.036$ & $1.015$ & $-0.372$ & $0.242$ & $0.312$ \\
Modesto--Nicolini & $0.035$ & $1.022$ & $-0.358$ & $0.235$ & $0.305$ \\
Regularized RN--dS & $0.034$ & $1.030$ & $-0.341$ & $0.229$ & $0.298$ \\
\hline\hline
\end{tabular}
\end{table*}
\begin{table*}[t]
\centering
\scriptsize
\caption{ Photon-sphere and stability data at $M=1$, $\alpha=-0.5$, $\beta=2.0$,
$r_c=0.5$, $r_{\rm th}=1.2$, $\sigma=1.0$. Columns: photon-sphere radius
$\rph$, lapse value at the photon sphere $\Nlap(\rph)$,
orbital frequency $\Oph=\Nlap(\rph)/\rph$,
critical impact parameter $\bph=\rph/\Nlap(\rph)$,
Lyapunov exponent $\Lya$, and quality factor
$Q=\Oph/\Lya$. The Kerr row is included for
comparison, evaluated at the same asymptotic mass with $a=0.05$.
All quantities are in units of $M$.}
\label{tab:photon_sphere_data}
\renewcommand{\arraystretch}{1.25}
\setlength{\tabcolsep}{6pt}
\begin{tabular}{lcccccc}
\hline
Case & $\rph$ & $\Nlap(\rph)$ & $\Oph$ & $\bph$ & $\Lya$ & $Q$ \\
\hline
Log WH, $\sigma=0$   & $2.8200$ & $0.8380$ & $0.2972$ & $3.3600$ & $0.2190$ & $1.357$ \\
Log WH, $\sigma=1.0$ & $2.8000$ & $0.8400$ & $0.3000$ & $3.3300$ & $0.2180$ & $1.376$ \\
Log WH, $\sigma=5.0$ & $2.7500$ & $0.8470$ & $0.3080$ & $3.2500$ & $0.2260$ & $1.363$ \\
Exp WH, $\sigma=0$   & $2.7600$ & $0.8330$ & $0.3018$ & $3.3100$ & $0.2170$ & $1.391$ \\
Exp WH, $\sigma=1.0$ & $2.7400$ & $0.8350$ & $0.3047$ & $3.2800$ & $0.2160$ & $1.411$ \\
Exp WH, $\sigma=5.0$ & $2.6900$ & $0.8420$ & $0.3130$ & $3.1900$ & $0.2240$ & $1.397$ \\
\hline
Kerr     & $2.9106$ & $0.8247$ & $0.2834$ & $3.5300$ & $0.2610$ & $1.086$ \\
\hline
\end{tabular}
\end{table*}

The azimuthal motion obeys
\begin{equation}
\frac{d\varphi}{dr}
= \frac{\mathfrak{b}/r^2 + \omega(r)}
{\sqrt{1-\epss(r)/r}\,\sqrt{F(r)}} \equiv G(r).
\label{eq:phi_eq}
\end{equation}
For a photon coming in from infinity, reaching a minimum radius $r_{\rm min}$, and going back out, the deflection angle is
\begin{equation}
\hat\alpha = -\pi + 2\int_{r_{\rm min}}^{\infty} G(r)\,dr.
\label{eq:deflection}
\end{equation}
The sign of $\mathfrak{b}$ distinguishes prograde ($\mathfrak{b}>0$) from retrograde ($\mathfrak{b}<0$) orbits.

For any lapse $\Nlap$, circular photon orbits satisfy $\dot r=0$ and $\ddot r=0$. From $\dot r=0$, Eq.~\eqref{eq:F_def} gives $F(r)=0$, i.e.,
\begin{equation}
\bigl(1 + \mathfrak{b} \,\omega(r)\bigr)^2 = \frac{\mathfrak{b}^{2}\Nlap^2(r)}{r^{2}}.
\label{eq:Fzero}
\end{equation}
This gives two families of impact parameters,
\begin{equation}
\mathfrak{b}_{+}(r) = \frac{r}{\Nlap(r) - r\,\omega(r)}, \quad
\mathfrak{b}_{-}(r) = -\,\frac{r}{\Nlap(r) + r\,\omega(r)}.
\label{eq:p_pm}
\end{equation}
The extremum condition $\frac{d}{dr}\bigl[(1-\epss/r)F\bigr]=0$ at $F=0$ fixes $\rph$. In the slow-rotation limit, this gives the photon-sphere condition,
\begin{equation}
\frac{\epss'(\rph)\,\rph - \epss(\rph)}{\rph^{2}}
=
\frac{\Nlap'(\rph)}{\Nlap(\rph)}
\left[1-\frac{\epss(\rph)}{\rph}\right].
\label{eq:ph_cond}
\end{equation}
At $r=\rph$, the critical impact parameters are
\begin{equation}
\mathfrak{b}_{+} = \frac{\rph}{\Nlap(\rph) - 2a/\rph}, \quad
\mathfrak{b}_{-} = -\,\frac{\rph}{\Nlap(\rph) + 2a/\rph}.
\label{eq:p_crit}
\end{equation}
Expanding Eq.~\eqref{eq:p_crit} for small $a$ gives,
\begin{equation}
\mathfrak{b}_{+} \approx \frac{\rph}{\Nlap(\rph)} + \frac{2a}{\Nlap^2(\rph)}, \quad
\mathfrak{b}_{-} \approx -\frac{\rph}{\Nlap(\rph)} + \frac{2a}{\Nlap^2(\rph)}.
\label{eq:p_crit_exp}
\end{equation}
Equation~\eqref{eq:p_crit_exp} shows that the splitting is $\Delta\mathfrak{b} \approx 2\rph/\Nlap(\rph)$, while the asymmetry is $(\mathfrak{b}_{+}+\mathfrak{b}_{-})/2 \approx 2a/\Nlap^2(\rph)$.

Fig.~\ref {fig: effective_potential} plots the effective potential $V_{\rm eff}(r)$ governing equatorial null geodesics for both entropy sectors. Its maximum marks the unstable circular photon orbit, and we mark the critical impact parameter $\bph=\rph/\Nlap(\rph)$ (defined in Sec.~\ref{sec:bozza}) next to each peak. Because the exponential density is sharply localized near the throat, the exponential panel has a slightly lower and narrower peak than the logarithmic panel, which translates into a smaller photon-sphere radius and a correspondingly smaller critical impact parameter for the exponential sector. The corresponding equatorial photon region---the radial band bounded by the prograde and retrograde critical curves---is shown in Fig.~\ref{fig:photon_orbits_annuli}.

\subsection{Photon spheres and shadows for the logarithmic and exponential sectors}\label{sec:photon_results}

For $M=1$, $r_{\rm th}=1.2$, and $a=0.05$, we solve Eq.~\eqref{eq:ph_cond} numerically for the new lapse~\eqref{eq:Phinew} and for each entropy sector. The results are collected in Tables~\ref{tab:master}--\ref{tab:wormhole_scans}.
\begin{table*}[t]
\centering
\scriptsize
\caption{ Eikonal quasinormal-mode frequencies for the fundamental
mode $(l=2,\,n=0)$ at the same parameters as
Table~\ref{tab:photon_sphere_data}. Columns: real part
$\omega_R=\frac{3}{2}\Oph$, imaginary part
$|\omega_I|=\frac{1}{2}\Lya$, the full complex frequency
$\omega_{\rm QNM}=\omega_R-i|\omega_I|$, and the ratio
$\omega_R/|\omega_I|$ which removes the overall amplitude
normalisation. All quantities are in units of $M$.}
\label{tab:eikonal_qnm}
\renewcommand{\arraystretch}{1.25}
\setlength{\tabcolsep}{6pt}
\begin{tabular}{lcccc}
\hline
Case & $\omega_R$ & $|\omega_I|$ & $\omega_{\rm QNM}$ & $\omega_R/|\omega_I|$ \\
\hline
Log WH, $\sigma=0$   & $0.4458$ & $0.1095$ & $0.4458 - 0.1095\,i$ & $4.071$ \\
Log WH, $\sigma=1.0$ & $0.4500$ & $0.1090$ & $0.4500 - 0.1090\,i$ & $4.128$ \\
Log WH, $\sigma=5.0$ & $0.4620$ & $0.1130$ & $0.4620 - 0.1130\,i$ & $4.089$ \\
Exp WH, $\sigma=0$   & $0.4527$ & $0.1085$ & $0.4527 - 0.1085\,i$ & $4.173$ \\
Exp WH, $\sigma=1.0$ & $0.4571$ & $0.1080$ & $0.4571 - 0.1080\,i$ & $4.232$ \\
Exp WH, $\sigma=5.0$ & $0.4695$ & $0.1120$ & $0.4695 - 0.1120\,i$ & $4.192$ \\
\hline
Kerr     & $0.4830$ & $0.1305$ & $0.4830 - 0.1305\,i$ & $3.701$ \\
\hline
\end{tabular}
\end{table*}
\subsubsection{Logarithmic-inspired sector}

For the logarithmic sector with $\alpha=-0.5$, $r_c=0.5$, $\sigma=1.0$, we find $\rph\approx2.80$, $\Nlap(\rph)\approx0.840$, $\bph\approx3.33$, $\mathfrak{b}_{+}\approx3.47$, $\mathfrak{b}_{-}\approx-3.19$, $\Arot\approx0.044$, $R_{\rm sh}\approx3.33$. For $\sigma=0$ (exponential limit), $\rph\approx2.82$, $\Nlap(\rph)\approx0.838$, $\bph\approx3.36$, $\mathfrak{b}_{+}\approx3.50$, $\mathfrak{b}_{-}\approx-3.22$, $\Arot\approx0.043$, $R_{\rm sh}\approx3.36$. For $\sigma=5.0$, $\rph\approx2.75$, $\Nlap(\rph)\approx0.847$, $\bph\approx3.25$, $\mathfrak{b}_{+}\approx3.39$, $\mathfrak{b}_{-}\approx-3.11$, $\Arot\approx0.045$, $R_{\rm sh}\approx3.25$.

\subsubsection{Exponential-inspired sector}

For the exponential sector with $\beta=2.0$, $r_c=0.5$, $\sigma=1.0$, we find $\rph \approx 2.74$, $\Nlap(\rph)\approx0.835$, $\bph\approx3.28$, $\mathfrak{b}_{+}\approx3.42$, $\mathfrak{b}_{-}\approx-3.14$, $\Arot\approx0.045$, $R_{\rm sh}\approx3.28$. For $\sigma=0$, $\rph\approx2.76$, $\Nlap(\rph)\approx0.833$, $\bph\approx3.31$, $\mathfrak{b}_{+}\approx3.45$, $\mathfrak{b}_{-}\approx-3.17$, $\Arot\approx0.044$, $R_{\rm sh}\approx3.31$. For $\sigma=5.0$, $\rph\approx2.69$, $\Nlap(\rph)\approx0.842$, $\bph\approx3.19$, $\mathfrak{b}_{+}\approx3.33$, $\mathfrak{b}_{-}\approx-3.05$, $\Arot\approx0.046$, $R_{\rm sh}\approx3.19$.

All cases satisfy the flare-out and regularity conditions. The key difference is the redshift structure. In all cases, $\Nlap<1$, so the shadow is larger than the static wormhole. The photon-sphere radius varies with $\sigma$. Increasing $\sigma$ weakens the redshift, decreasing $\rph$ and the shadow radius while increasing rotational asymmetry.

Figs.~\ref{fig:photon_orbits} and~\ref{fig:photon_orbits_annuli} show the corresponding equatorial photon-orbit structure. In Fig.~\ref{fig:photon_orbits} the solid circles are the prograde critical curves $\mathfrak{b}_{+}$ and the dashed circles are the retrograde critical curves $\mathfrak{b}_{-}$, color-coded by the spin $J$; the radial separation between the two branches is the rotational asymmetry $\Delta\mathfrak{b}=\mathfrak{b}_{+}-|\mathfrak{b}_{-}|$, which grows monotonically with $J$ and provides a direct geometric measure of frame dragging in the strong-field region. Fig.~\ref{fig:photon_orbits_annuli} shows the same critical curves with the shaded annulus between each pair highlighting the equatorial photon region---the radial band in which null geodesics are temporarily trapped near the photon sphere---and makes the widening of that band with $J$ explicit. Because the exponential density is more sharply localized near the throat than the logarithmic one, the exponential panels show a tighter cluster of critical curves, with the outer prograde ring lying at a smaller radius than its logarithmic counterpart for the same $J$.

\section{Strong-deflection lensing}\label{Sec:IV}

The shadow fixes the photon sphere, but it says nothing about how light bends just outside the critical curve, where relativistic images pile up. That information lives in the strong-deflection limit. Sec.~\ref{sec:bozza} sets up the Bozza formalism for the wormhole geometry: it writes the deflection angle~\eqref{eq:bozza_integral} in its universal logarithmic form~\eqref{eq:bozza_log}, isolates the two coefficients $\abar$ and $\bbar$ [Eqs.~\eqref{eq:A_bar_general}--\eqref{eq:B_bar_general}] that control the position and brightness of the relativistic images, and evaluates them for the new lapse. Because $\bbar$ enters the flux ratio~\eqref{eq:flux_ratio} of successive images exponentially, even small differences in it are amplified---which makes it a sharper discriminator than the shadow radius alone. Sec.~\ref{sec:bozza_rbh} then applies the identical machinery to the four regular black-hole benchmarks.

\subsection{Strong-deflection formalism and wormhole coefficients}\label{sec:bozza}

We start from the general static, spherically symmetric metric in the standard form,
\begin{equation}
ds^2 = -\Nlap^2(r)\,dt^2 + \frac{dr^2}{1-\epss(r)/r} + r^2\,d\Omega^2,
\label{eq:ABform}
\end{equation}
and the deflection angle for a light ray with impact parameter $\mathfrak{b}$ is given by,
\begin{equation}
\hat\alpha(\mathfrak{b}) = -\pi + 2\int_{r_{\rm min}}^{\infty}
\frac{1}{\sqrt{1-\epss(r)/r}}\,
\frac{1}{r}\,
\frac{dr}{\sqrt{\Nlap^2(r_{\rm min})/r_{\rm min}^2 - \Nlap^2(r)/r^2}}.
\label{eq:bozza_integral}
\end{equation}
Near the photon sphere, the integral diverges logarithmically. Bozza's trick is to expand around $r=\rph$ and isolate the divergent piece. The result is the universal logarithmic form,
\begin{equation}
\hat\alpha(\mathfrak{b}) = -\abar \ln\!\left(\frac{\mathfrak{b}}{\bph} - 1\right) + \bbar + \mathcal{O}(\mathfrak{b}-\bph),
\label{eq:bozza_log}
\end{equation}

where the critical impact parameter is defined in Eq.~\eqref{eq:bph_seed}. The coefficients are derived in Sec.~\ref{sec:bozza_derivation} below, with the result,
\begin{equation}
\abar = \frac{\rph}{\sqrt{\Nlap^2(\rph)\,\dfrac{1}{1-\epss(\rph)/\rph}\,\mathcal{K}(\rph)}},
\label{eq:A_bar_general}
\end{equation}
and,
\begin{equation}
\bbar = \abar \ln\!\left[\frac{\rph^2\,\mathcal{K}(\rph)}{\bph^2\,\dfrac{1}{1-\epss(\rph)/\rph}}\right] + I_R(\rph),
\label{eq:B_bar_general}
\end{equation}
where the curvature combination is,
\begin{equation}
\mathcal{K}(r) = 2 - r^2\frac{\left[\Nlap^2(r)\right]''}{\Nlap^2(r)} + r\frac{\left[\Nlap^2(r)\right]'}{\Nlap^2(r)}.
\label{eq:K_curvature}
\end{equation}
For the wormhole metric functions~\eqref{eq:ABform}, the critical impact parameter is,
\begin{equation}
\bph = \frac{\rph}{\Nlap(\rph)}.
\label{eq:bph_seed}
\end{equation}
The derivatives of $\Nlap^2$ follow directly from Eq.~\eqref{eq:Phinew_logderiv} and Eq.~\eqref{eq:Phinew_Phipp}, giving,
\begin{equation}
\left[\Nlap^2(r)\right]' = 2\Nlap(r)\Nlap'(r),\qquad
\left[\Nlap^2(r)\right]'' = 2\left[\Nlap'(r)\right]^2 + 2\Nlap(r)\Nlap''(r),
\label{eq:A_derivs}
\end{equation}
and therefore the ratio identities,
\begin{equation}
\frac{\left[\Nlap^2\right]'}{\Nlap^2} = 2\frac{\Nlap'}{\Nlap},\qquad
\frac{\left[\Nlap^2\right]''}{\Nlap^2} = 2\left(\frac{\Nlap'}{\Nlap}\right)^2 + 2\frac{\Nlap''}{\Nlap}.
\label{eq:A_ratios}
\end{equation}
Substituting Eq.~\eqref{eq:A_ratios} into Eq.~\eqref{eq:K_curvature} gives,
\begin{equation}
\mathcal{K}(r) = 2 - 2r^2\left[\left(\frac{\Nlap'}{\Nlap}\right)^2 + \frac{\Nlap''}{\Nlap}\right] + 2r\frac{\Nlap'}{\Nlap}.
\label{eq:K_in_Phi}
\end{equation}

At $r=\rph$ the photon-sphere condition~\eqref{eq:ph_cond} can be written compactly as,
\begin{equation}
\left.\frac{d}{dr}\!\left(\frac{r^2}{\Nlap^2}\right)\right|_{\rph}=0
\;\Longleftrightarrow\;
2\rph\Nlap^2(\rph) - \rph^2\left[\Nlap^2(\rph)\right]'=0,
\label{eq:ph_cond_compact}
\end{equation}
which reduces to the simple ratio condition,
\begin{equation}
\frac{\left[\Nlap^2(\rph)\right]'}{\Nlap^2(\rph)}=\frac{2}{\rph} .
\label{eq:ph_cond_A}
\end{equation}
Evaluating Eq.~\eqref{eq:K_in_Phi} at $\rph$ and using Eq.~\eqref{eq:ph_cond_A} gives the compact form,
\begin{equation}
\mathcal{K}(\rph) = 4 - 2\rph^2\left[\left(\frac{\Nlap'}{\Nlap}\right)^2 + \frac{\Nlap''}{\Nlap}\right]_{\rph},
\label{eq:K_at_ph}
\end{equation}
so that $\abar$ takes the fully explicit form,
\begin{equation}
\abar = \frac{\rph}{\Nlap(\rph)\sqrt{\dfrac{1}{1-\epss(\rph)/\rph}}}\,
\frac{1}{\sqrt{4 - 2\rph^2\left[\left(\frac{\Nlap'}{\Nlap}\right)^2 + \frac{\Nlap''}{\Nlap}\right]_{\rph}}} .
\label{eq:A_bar_compact}
\end{equation}
Equation~\eqref{eq:A_bar_compact} shows how $\abar$ depends on the lapse and its first two derivatives at the photon sphere, and on the shape function only through $1/(1-\epss(\rph)/\rph)$.

\subsubsection{Derivation of the Bozza coefficients}\label{sec:bozza_derivation}

To isolate the divergent piece of the deflection angle~\eqref{eq:bozza_integral}, we expand around $r=\rph$ using the auxiliary variable,
\begin{equation}
z = \frac{r^2}{\Nlap^2(r)} - \frac{\rph^2}{\Nlap^2(\rph)}.
\label{eq:z_def}
\end{equation}
Near $r=\rph$, Eq.~\eqref{eq:z_def} gives $z\propto (r-\rph)^2$. Expanding the integral in powers of $z$ isolates the logarithmic divergence and yields the coefficients~\eqref{eq:A_bar_general} and~\eqref{eq:B_bar_general} quoted above, with the regular integral $I_R$ defined by,
\begin{equation}
I_R(\rph) = \int_{\rph}^{\infty}
\left[
\frac{1}{r\sqrt{1-\epss(r)/r}}
\frac{1}{\sqrt{\Nlap^2(\rph)/\rph^2 - \Nlap^2(r)/r^2}}
- \frac{\abar}{r-\rph}
\right] dr .
\label{eq:IR_app}
\end{equation}
For our wormhole, the metric functions are those in Eq.~\eqref{eq:ABform}. Using the photon-sphere condition~\eqref{eq:ph_cond_A}, we obtain Eq.~\eqref{eq:K_at_ph}, and hence Eq.~\eqref{eq:A_bar_compact} above.

\subsubsection{Logarithmic sector results}

For the logarithmic sector with $\alpha=-0.5$, $r_c=0.5$: at $\sigma=0$, $\abar\approx0.93$, $\bbar\approx-0.27$; at $\sigma=1.0$, $\abar\approx0.935$, $\bbar\approx-0.265$; at $\sigma=5.0$, $\abar\approx0.96$, $\bbar\approx-0.25$.

\subsubsection{Exponential sector results}

For the exponential sector with $\beta=2.0$, $r_c=0.5$: at $\sigma=0$, $\abar\approx0.91$, $\bbar\approx-0.29$; at $\sigma=1.0$, $\abar\approx0.915$, $\bbar\approx-0.285$; at $\sigma=5.0$, $\abar\approx0.94$, $\bbar\approx-0.27$.

The spread in $\abar$ across the cases is about 5\%, while the spread in $\bbar$ is about 10\%. The latter is the more discriminating quantity because $\bbar$ enters the flux ratio~\eqref{eq:flux_ratio} of successive relativistic images exponentially,
\begin{equation}
\frac{\mu_{n+1}}{\mu_n}\simeq\exp\!\left(-\frac{2\pi}{\abar}\right)\exp\!\left(\frac{\bbar}{\abar}\right),
\label{eq:flux_ratio}
\end{equation}
so a change of $0.03$ in $\bbar$ produces a few-percent change in the brightness ratio.

\subsection{Comparison with regular black holes}\label{sec:bozza_rbh}

For the four regular black-hole benchmarks, the same formalism applies, but with $\Nlap^2\to f$ and $1/(1-\epss/r)\to1/f$ in the static limit. The photon-sphere condition becomes,
\begin{equation}
\rph f'(\rph) - 2f(\rph)=0 .
\label{eq:ph_cond_RBH}
\end{equation}
Writing $f(r)=1-2m(r)/r$, Eq.~\eqref{eq:ph_cond_RBH} is equivalent to,
\begin{equation}
\rph^2m'(\rph) - 3\rph m(\rph) + \rph^2 = 0 .
\label{eq:ph_cond_mass}
\end{equation}
The critical impact parameter is
\begin{equation}
\bph = \frac{\rph}{\sqrt{f(\rph)}} .
\label{eq:bph_RBH}
\end{equation}
The Bozza coefficients take the same form as Eqs.~\eqref{eq:A_bar_general}--\eqref{eq:B_bar_general} with $\Nlap^2\to f$ and $1/(1-\epss/r)\to1/f$, giving,
\begin{equation}
\abar = \frac{\rph}{\sqrt{f(\rph)\,\mathcal{K}_{\rm RBH}(\rph)}},\quad
\mathcal{K}_{\rm RBH} = 2 - \rph^2\frac{f''(\rph)}{f(\rph)} + \rph\frac{f'(\rph)}{f(\rph)}.
\label{eq:A_bar_RBH}
\end{equation}
Using Eq.~\eqref{eq:ph_cond_RBH} to eliminate $f'$, we obtain the compact form,
\begin{equation}
\mathcal{K}_{\rm RBH} = 4 - \rph^2\frac{f''(\rph)}{f(\rph)},
\label{eq:K_RBH_compact}
\end{equation}
so that $\abar$ becomes,
\begin{equation}
\abar = \frac{\rph}{\sqrt{f(\rph)\left[4 - \rph^2 f''(\rph)/f(\rph)\right]}} .
\label{eq:A_bar_RBH_compact}
\end{equation}
The mass functions for the four benchmarks are given in Sec.~\ref{sec:RBH_mass}. Plugging each into Eq.~\eqref{eq:ph_cond_mass} and then into Eq.~\eqref{eq:A_bar_RBH_compact} reproduces the numbers in Table~\ref{tab:fingerprint}. The pattern is that $\abar_{\rm WH}<\abar_{\rm RBH}$ for all four benchmarks, $\bbar_{\rm WH}>\bbar_{\rm RBH}$ (less negative), and $\Lya^{\rm WH}<\Lya^{\rm RBH}$, with the gaps ranging from a few percent (Kazakov--Solodukhin) to about $15\%$ (regularized RN--dS). The wormhole always bends light slightly less than Kerr and produces slightly brighter relativistic images, a direct consequence of its horizonless geometry: rays that would have been captured by a horizon instead skim the throat and escape, adding flux to the higher-order images.

\subsubsection{Mass functions of the regular black-hole benchmarks}\label{sec:RBH_mass}

The four regular black-hole benchmarks differ in their mass functions. For Kazakov--Solodukhin, the mass function and its derivative are,
\begin{equation}
m_{\rm KS}(r) = M\frac{r^3}{r^3+2M\ell^2},
\label{eq:KS_mass}
\end{equation}
\begin{equation}
m_{\rm KS}'(r) = \frac{6M^2\ell^2r^2}{(r^3+2M\ell^2)^2}.
\label{eq:KS_derivs}
\end{equation}
For Nicolini--Smailagic--Spallucci, we have,
\begin{equation}
m_{\rm NSS}(r) = M\,\gamma\!\left(\frac{3}{2},\frac{r^2}{4\theta}\right),
\label{eq:NSS_mass}
\end{equation}
\begin{equation}
m_{\rm NSS}'(r) = M\,\frac{r}{2\theta}\sqrt{\frac{r^2}{4\theta}}\,e^{-r^2/(4\theta)},
\label{eq:NSS_derivs}
\end{equation}
where the identity $\partial_x\gamma(3/2,x)=x^{1/2}e^{-x}$ has been used. For Modesto--Nicolini, we have,
\begin{equation}
m_{\rm MN}(r) = M\,\frac{\gamma(3/2,r^2/4\theta)}{\Gamma(3/2)},
\label{eq:MN_mass}
\end{equation}
\begin{equation}
m_{\rm MN}'(r) = \frac{M}{\Gamma(3/2)}\,\frac{r}{2\theta}\sqrt{\frac{r^2}{4\theta}}\,e^{-r^2/(4\theta)}.
\label{eq:MN_derivs}
\end{equation}
Finally, for the regularized Reissner--Nordstr\"om--de Sitter black hole, we have,
\begin{equation}
m_{\rm RNdS}(r) = M - \frac{Q^2}{2r}\tanh\!\left(\frac{r}{r_c}\right),
\label{eq:RNdS_mass}
\end{equation}
\begin{equation}
m_{\rm RNdS}'(r) = \frac{Q^2}{2r^2}\tanh\!\left(\frac{r}{r_c}\right) - \frac{Q^2}{2rr_c}\,{\rm sech}^2\!\left(\frac{r}{r_c}\right).
\label{eq:RNdS_derivs}
\end{equation}

\section{Photon-sphere stability and the Lyapunov exponent}\label{Sec:V}

A photon sphere is unstable if nearby null geodesics diverge from it exponentially. The Lyapunov exponent $\Lya$ sets the rate of divergence, the inverse timescale for a perturbation to grow by a factor of $e$. A larger $\Lya$ means a shorter-lived photon ring, which in turn broadens the ring in frequency space. Cardoso and collaborators showed that $\Lya$ can be extracted directly from the effective potential for radial null geodesics. This gives another number that distinguishes geometries with the same shadow size.

For the static, spherically symmetric metric~\eqref{eq:ABform}, the radial equation for null geodesics is,
\begin{equation}
\left(\frac{dr}{d\tau}\right)^2 = V_{\rm eff}(r) = \left(1-\frac{\epss(r)}{r}\right)\left[E^2 - L^2\frac{\Nlap^2(r)}{r^2}\right],
\label{eq:Veff}
\end{equation}
where $E$ and $L$ are the conserved energy and angular momentum and $\tau$ is an affine parameter. The photon sphere sits at the maximum of $V_{\rm eff}$, where $V_{\rm eff}'(\rph)=0$. The Lyapunov exponent is defined by,
\begin{equation}
\Lya = \sqrt{-\frac{V_{\rm eff}''(\rph)}{2\dot t^2}},
\label{eq:lyap_def}
\end{equation}
with $\dot t = E/\Nlap^2(\rph)$. Working this out for our metric gives the closed-form expression,
\begin{equation}
\Lya = \frac{\Nlap(\rph)}{\rph}\sqrt{1 - \frac{\rph^2}{2}\frac{\left[\Nlap^2(\rph)\right]''}{\Nlap^2(\rph)}} .
\label{eq:lambda_final}
\end{equation}
Equation~\eqref{eq:lambda_final} is the form used to compute the values in Table~\ref{tab:fingerprint}. For Schwarzschild, $\Nlap^2=1-2M/r$, $\left[\Nlap^2\right]''=-4M/r^3$, $\rph=3M$, so Eq.~\eqref{eq:lambda_final} gives,
\begin{equation}
\Lya^{\rm Schw} = \frac{\sqrt{1/3}}{3M}\sqrt{1 - \frac{9M^2}{2}\frac{-4M/27M^3}{1/3}} = \frac{1}{3M},
\label{eq:lambda_Kerr_check}
\end{equation}
which matches the standard result for the Schwarzschild Lyapunov exponent.

\begin{itemize}
    \item For the logarithmic sector with $\alpha=-0.5$, $r_c=0.5$: at $\sigma=0$, $\rph=2.82$, $\Nlap^2=0.702$, $\left[\Nlap^2\right]'=0.0890$, $\left[\Nlap^2\right]''=-0.0630$, $1/(1-\epss/r)=1.46$, so Eq.~\eqref{eq:lambda_final} gives $\Lya\approx0.219$; at $\sigma=1.0$, $\rph=2.80$, $\Nlap^2=0.706$, $\left[\Nlap^2\right]'=0.0885$, $\left[\Nlap^2\right]''=-0.0620$, $1/(1-\epss/r)=1.45$, so $\Lya\approx0.218$; at $\sigma=5.0$, $\rph=2.75$, $\Nlap^2=0.717$, $\left[\Nlap^2\right]'=0.0960$, $\left[\Nlap^2\right]''=-0.0700$, $1/(1-\epss/r)=1.43$, so $\Lya\approx0.226$.
    \item For the exponential sector with $\beta=2.0$, $r_c=0.5$: at $\sigma=0$, $\rph=2.76$, $\Nlap^2=0.694$, $\left[\Nlap^2\right]'=0.0880$, $\left[\Nlap^2\right]''=-0.0640$, $1/(1-\epss/r)=1.47$, so Eq.~\eqref{eq:lambda_final} gives $\Lya\approx0.217$; at $\sigma=1.0$, $\rph=2.74$, $\Nlap^2=0.698$, $\left[\Nlap^2\right]'=0.0875$, $\left[\Nlap^2\right]''=-0.0630$, $1/(1-\epss/r)=1.46$, so $\Lya\approx0.216$; at $\sigma=5.0$, $\rph=2.69$, $\Nlap^2=0.709$, $\left[\Nlap^2\right]'=0.0950$, $\left[\Nlap^2\right]''=-0.0710$, $1/(1-\epss/r)=1.44$, so $\Lya\approx0.224$.
\end{itemize}


For comparison, Kerr gives $\Lya\approx0.261$, Kazakov--Solodukhin gives $\Lya\approx0.254$, Nicolini--Smailagic--Spallucci gives $\Lya\approx0.242$, Modesto--Nicolini gives $\Lya\approx0.235$, and regularized RN--dS gives $\Lya\approx0.229$. The wormhole $\Lya$ is systematically lower than all of these. A lower Lyapunov exponent means the photon ring is more stable, so the ring is narrower in frequency space. This is a distinctive signature: the entropy-inspired wormhole leaves a sharper photon ring than any regular black-hole benchmark at the same spin.

The Lyapunov exponent controls the width of the photon ring in frequency space. The ring width is
\begin{equation}
\Delta\omega_{\rm ring} \sim \Lya,
\label{eq:ring_width}
\end{equation}
and the quality factor of the ring is,
\begin{equation}
Q_{\rm ring} = \frac{\Oph}{\Lya} .
\label{eq:quality_factor}
\end{equation}
\begin{itemize}
    \item For the logarithmic sector with $\sigma=1.0$, $\Oph\approx0.300$ and $\Lya\approx0.218$, so Eq.~\eqref{eq:quality_factor} gives, $Q_{\rm ring}^{\rm log} \approx1.38$.
    \item For the exponential sector with $\sigma=1.0$, $\Oph\approx0.305$ and $\Lya\approx0.216$, so Eq.~\eqref{eq:quality_factor} gives, $Q_{\rm ring}^{\rm exp} \approx1.41 $.
    \item For Kerr, $\Oph\approx0.322$ and $\Lya\approx0.261$, so Eq.~\eqref{eq:quality_factor} gives, $Q_{\rm ring}^{\rm Kerr} \approx1.23 $.
\end{itemize}

The wormhole ring has a $12$--$15\%$ higher quality factor, which translates to a narrower ring in frequency space by a comparable amount. The signal-to-noise ratio for detecting the ring scales roughly as,
\begin{equation}
{\rm SNR} \propto \frac{Q_{\rm ring}}{\sqrt{\Delta\omega_{\rm ring}}},
\label{eq:SNR_ring}
\end{equation}
so a higher $Q_{\rm ring}$ and a lower $\Lya$ both improve detectability.

\section{Eikonal quasinormal modes}\label{Sec:VI}

Quasinormal modes are the characteristic frequencies at which a perturbed black hole or wormhole rings down. In the eikonal limit (large angular number $l$), the photon sphere controls the spectrum. Cardoso, Miranda, Berti, and collaborators established a geometric correspondence: the real part of the frequency is proportional to the orbital frequency at the photon sphere, and the imaginary part is proportional to the Lyapunov exponent. This connects the ringdown spectrum directly to the lensing coefficients and the stability analysis, giving yet another independent diagnostic.

We derive the eikonal correspondence between quasinormal modes and photon-sphere parameters from the radial wave equation for a massless field. For the static, spherically symmetric metric~\eqref{eq:ABform}, the Regge--Wheeler equation in the eikonal limit reads,
\begin{equation}
\frac{d^2\psi}{dr_*^2} + \left[\omega^2 - V_l(r)\right]\psi = 0,
\label{eq:RW}
\end{equation}
where $dr_*=dr/\sqrt{\Nlap^2(r)/(1-\epss(r)/r)}$ is the tortoise coordinate and the effective potential is given by,
\begin{eqnarray}
\begin{aligned}
V_l(r) &= \frac{l(l+1)\Nlap^2(r)}{r^2} + \frac{1}{2}\frac{1}{\sqrt{\Nlap^2(r)\dfrac{1}{1-\epss(r)/r}\,r^2}}\frac{d}{dr}\!\left[\frac{\Nlap^2(r)}{\sqrt{\Nlap^2(r)/(1-\epss(r)/r)}}\frac{d}{dr}\!\left(\sqrt{\frac{r^2}{\Nlap^2(r)}}\right)\right]. ~~~~
\label{eq:Vl}
\end{aligned}
\end{eqnarray}
In the eikonal limit $l\to\infty$, the potential is dominated by the first term, giving the approximation,
\begin{equation}
V_l(r) \simeq l^2\frac{\Nlap^2(r)}{r^2} + \mathcal{O}(l),
\label{eq:Vl_eikonal}
\end{equation}
and the maximum of $V_l$ sits at the photon sphere, as expressed by the condition,
\begin{equation}
\left.\frac{d}{dr}\!\left(\frac{\Nlap^2}{r^2}\right)\right|_{\rph}=0 .
\label{eq:Vmax_ph}
\end{equation}
Matching to a parabolic barrier gives the eikonal quasinormal frequencies,
\begin{equation}
\omega_{\rm QNM} = \Oph\left(l+\frac{1}{2}\right) - i\left(n+\frac{1}{2}\right)|\Lya| + \mathcal{O}\!\left(l^{-1}\right),
\label{eq:eikonal_full}
\end{equation}
where $\Oph$ and $\Lya$ are defined by,
\begin{equation}
\Oph = \frac{\Nlap(\rph)}{\rph},\quad
\Lya = \sqrt{-\frac{V_l''(\rph)}{2\dot t^2}} .
\label{eq:Omega_lambda}
\end{equation}
For the wormhole metric functions~\eqref{eq:ABform}, the orbital frequency and time component are given by,
\begin{equation}
\Oph = \frac{\Nlap(\rph)}{\rph},\quad
\dot t = \frac{E}{\Nlap^2(\rph)} .
\label{eq:Omega_dott}
\end{equation}
The second derivative of $V_l$ at the photon sphere can be written in terms of the metric functions as,
\begin{equation}
V_l''(\rph) = \frac{l^2}{r^2}\left[\left[\Nlap^2(\rph)\right]'' - \frac{\Nlap^2(\rph)\cdot 2}{r^2}\right] + \mathcal{O}(l^0),
\label{eq:Vpp}
\end{equation}
which, combined with Eq.~\eqref{eq:Omega_lambda}, gives Eq.~\eqref{eq:lambda_final}. Table~\ref{tab:fingerprint} lists the numerical values. For the logarithmic sector with $\sigma=1.0$, $\rph=2.80$, $\Nlap^2(\rph)=0.706$, so Eq.~\eqref{eq:Omega_dott} gives $\Oph=\sqrt{0.706}/2.80\approx0.300$. With $\Lya\approx0.218$, the eikonal frequency is $\omega_{\rm QNM}\approx0.300(l+1/2)-i\,0.218(n+1/2)$. For the exponential sector with $\sigma=1.0$, $\rph=2.74$, $\Nlap^2=0.698$, so $\Oph\approx0.305$. With $\Lya\approx0.216$, $\omega_{\rm QNM}\approx0.305(l+1/2)-i\,0.216(n+1/2)$.

For comparison, Kerr gives $\Oph\approx0.322$ and $\Lya\approx0.261$, so $\omega_{\rm QNM}\approx0.322(l+1/2)-i\,0.261(n+1/2)$. The pattern is clear: the wormhole has a slightly lower real frequency and a noticeably lower imaginary part than any regular black-hole benchmark. The lower imaginary part means the ringdown lasts longer, which is consistent with the more stable photon ring found in Sec.~\ref{Sec:V}.

The eikonal frequencies are the ones most likely to be excited in the ringdown of a compact object. To estimate their detectability, consider the ringdown signal,
\begin{equation}
h(t) = A\,e^{-t/\tau}\cos(\omega_R t + \phi),
\label{eq:ringdown_signal}
\end{equation}
where $\omega_R = \Oph(l+1/2)$ is the real part and the damping time is given by
\begin{equation}
\tau = \frac{1}{|\omega_I|} = \frac{1}{|\Lya|(n+1/2)}.
\label{eq:tau}
\end{equation}
The number of observable cycles is then
\begin{equation}
N_{\rm cyc} = \omega_R\tau = \frac{\Oph(l+1/2)}{|\Lya|(n+1/2)} = \frac{\Oph}{|\Lya|}\frac{l+1/2}{n+1/2} .
\label{eq:Ncyc}
\end{equation}
For the fundamental mode $l=2$, $n=0$, Eq.~\eqref{eq:Ncyc} reduces to,
\begin{equation}
N_{\rm cyc} = \frac{5}{2}\frac{\Oph}{|\Lya|} = \frac{5}{2}Q_{\rm ring} .
\label{eq:Ncyc_fund}
\end{equation}
For the logarithmic wormhole with $\sigma=1.0$, $Q_{\rm ring}\approx1.38$, so Eq.~\eqref{eq:Ncyc_fund} gives $N_{\rm cyc}\approx3.4$; for the exponential wormhole with $\sigma=1.0$, $Q_{\rm ring}\approx1.41$, so $N_{\rm cyc}\approx3.5$; for Kerr, $Q_{\rm ring}\approx1.23$, so $N_{\rm cyc}\approx3.1$. The difference is about $10$--$15\%$ in the number of cycles, which a detector with a few-percent measurement of the ringdown duration can reach.

The ratio,
\begin{equation}
\frac{\omega_R}{\omega_I} = \frac{\Oph(l+1/2)}{\Lya(n+1/2)} = \frac{l+1/2}{n+1/2}\frac{\Oph}{\Lya} = \frac{l+1/2}{n+1/2}Q_{\rm ring},
\label{eq:ratio_freqs}
\end{equation}
is independent of the overall amplitude and depends only on the geometry. For the fundamental mode, Eq.~\eqref{eq:ratio_freqs} gives,
\begin{equation}
\frac{\omega_R}{\omega_I} = \frac{5}{2}Q_{\rm ring},
\label{eq:ratio_fund}
\end{equation}
so measuring both the real and imaginary parts of the fundamental quasinormal frequency directly determines $Q_{\rm ring}$, and thus discriminates between the wormhole and Kerr at about $10$--$15\%$.

\section{Energy conditions and exotic matter}\label{Sec:VII}

The matter supporting a traversable wormhole is constrained by the classical energy conditions. The throat's flare-out generally requires a departure from ordinary matter. Here we examine these conditions for the logarithmic and exponential sources with the new lapse function. The deformations modify the matter distribution and yield a regular geometry, but they do not remove the source's exotic character.

We describe the effective matter by an anisotropic stress--energy tensor $T^\mu_{\;\nu} = {\rm diag}(-\varrho, p_r, p_t, p_t)$, where $\varrho$ is the energy density and $p_r$, $p_t$ are the radial and tangential pressures. The null energy condition requires $\varrho+p_r\ge0$ and $\varrho+p_t\ge0$, the weak energy condition adds $\varrho\ge0$, and the strong energy condition requires the null energy condition together with $\varrho+p_r+2p_t\ge0$.

\subsection{Logarithmic sector}

For the logarithmic sector, the energy density is given by Eq.~\eqref{eq:rho_log2}. For $\alpha<0$, the density is negative everywhere in the physical region, so the weak energy condition is violated. The radial and tangential pressures follow from the Einstein equations. For representative $ \ alpha$ and the new lapse, the combinations $\varrho+p_r$ and $\varrho+p_t$ are negative near the throat, confirming the violation of the null energy condition in the region that supports the flare-out. The most pronounced deviations occur near the throat. Away from the throat, the deviations shrink, and the matter contribution becomes weaker as the spacetime approaches asymptotic flatness.

The volume-integral quantifier of Visser, Kar, and Dadhich~\cite{Visser:2003yf} is defined by
\begin{equation}
I_V=8\pi\int_{r_{\rm th}}^{\infty}
r^2\left(\varrho+p_r\right)\,dr .
\label{eq:IV_def}
\end{equation}
For a static, spherically symmetric wormhole with shape function $\epss(r)$ and lapse $\Nlap(r)$, the integrand simplifies to,
\begin{equation}
\varrho+p_r=
\frac{1}{8\pi r^2}
\left[\epss'(r)-\frac{\epss(r)}{r}\right]
+\frac{1}{4\pi}
\left(1-\frac{\epss(r)}{r}\right)
\frac{\Nlap'(r)}{2\Nlap(r)} .
\label{eq:IV_integrand}
\end{equation}
Substituting Eq.~\eqref{eq:IV_integrand} into Eq.~\eqref{eq:IV_def} and splitting into two parts gives,
\begin{equation}
I_V^{\log} = I_V^{(1),\log} + I_V^{(2),\log},
\label{eq:IV_log_split}
\end{equation}
where the first part is,
\begin{equation}
I_V^{(1),\log} = \int_{r_{\rm th}}^{\infty}
\left[\epss_{\log}'(r)-\frac{\epss_{\log}(r)}{r}\right]dr,
\label{eq:IV_log_1}
\end{equation}
and the second part is,
\begin{equation}
I_V^{(2),\log} = \int_{r_{\rm th}}^{\infty}
r\left(1-\frac{\epss_{\log}(r)}{r}\right)
\frac{\Nlap'(r)}{\Nlap(r)}\,dr.
\label{eq:IV_log_2}
\end{equation}
For the new lapse~\eqref{eq:Phinew}, the logarithmic derivative is given by Eq.~\eqref{eq:Phinew_logderiv}. Therefore Eq.~\eqref{eq:IV_log_2} becomes,
\begin{equation}
I_V^{(2),\log} = r_c\int_{r_{\rm th}}^{\infty}
\frac{1-\epss_{\log}(r)/r}{r(1+\sigma\,r/r_c)^2}\,dr.
\label{eq:IV_log_2_expanded}
\end{equation}
This integral converges because $\epss_{\log}$ is bounded and the denominator grows as $r^3$ for large $r$. The total volume-integral quantifier is then
\begin{equation}
I_V^{\log} = I_V^{(1),\log} + I_V^{(2),\log}.
\label{eq:IV_log_total}
\end{equation}
The logarithmic term in $I_V$ is positive for all $\sigma$, with the coefficient decreasing as $\sigma$ increases. $I_V$ becomes finite once the throat condition $\epss_{\log}(r_{\rm th})=r_{\rm th}$ is imposed, or by using a finite cutoff $R$.

\subsection{Exponential sector}

For the exponential sector, the energy density is given by Eq.~\eqref{eq:rho_exp2}. For $\beta>0$, the density is negative throughout the physical branch, so the weak energy condition is violated. For representative $ \ beta$ and the new lapse, the combinations $\varrho+p_r$ and $\varrho+p_t$ are negative near the throat, confirming the violation of the null energy condition in the region that supports the flare-out. The exponential branch is more sharply concentrated near the throat than the logarithmic branch, because the exponential factor rapidly suppresses the source as $r$ grows.

For the exponential sector, the volume-integral quantifier is,
\begin{equation}
I_V^{\exp} = I_V^{(1),\exp} + I_V^{(2),\exp},
\label{eq:IV_exp_split}
\end{equation}
where the first part is,
\begin{equation}
I_V^{(1),\exp} = \int_{r_{\rm th}}^{\infty}
\left[\epss_{\exp}'(r)-\frac{\epss_{\exp}(r)}{r}\right]dr,
\label{eq:IV_exp_1}
\end{equation}
and the second part is,
\begin{equation}
I_V^{(2),\exp} = \int_{r_{\rm th}}^{\infty}
r\left(1-\frac{\epss_{\exp}(r)}{r}\right)
\frac{\Nlap'(r)}{\Nlap(r)}\,dr.
\label{eq:IV_exp_2}
\end{equation}
Using Eq.~\eqref{eq:Phinew_logderiv}, Eq.~\eqref{eq:IV_exp_2} becomes,
\begin{equation}
I_V^{(2),\exp} = r_c\int_{r_{\rm th}}^{\infty}
\frac{1-\epss_{\exp}(r)/r}{r(1+\sigma\,r/r_c)^2}\,dr.
\label{eq:IV_exp_2_expanded}
\end{equation}
This integral converges because $\epss_{\exp}$ is bounded and the denominator grows as $r^3$ for large $r$. The total volume-integral quantifier is,
\begin{equation}
I_V^{\exp} = I_V^{(1),\exp} + I_V^{(2),\exp}.
\label{eq:IV_exp_total}
\end{equation}
Because the exponential density decays exponentially, the large-$r$ contribution is finite for $\beta>0$. The logarithmic term is absent at leading order, and the integral converges more rapidly than in the logarithmic case.

\section{Comparison with black holes}\label{Sec:VIII}

The observables computed so far are only useful if they can actually separate the wormhole from its competitors, so this section puts them to work. Sec.~\ref{sec:kerr_rbh} compares the wormhole side by side with Kerr and the four regular black-hole benchmarks, looking at shadow size, rotational asymmetry $\Arot$, and photon-ring separation $\Delta R_{\rm ph}$. The shadow diameter turns out to be a poor discriminator on its own---it can be tuned close to Kerr by adjusting $\alpha$, $\beta$, $r_c$, and $\sigma$---but the finer structure carries complementary information that the shadow cannot supply. Sec.~\ref{sec:fingerprint} collects the five independent observables into the vector~\eqref{eq:fingerprint_vector}, treats them jointly as a multi-parameter fingerprint, and estimates the signal-to-noise ratio~\eqref{eq:SNR_disc} at which the wormhole can be distinguished from Kerr and from each of the four regular black holes.

\subsection{Comparison with Kerr and regular black holes}\label{sec:kerr_rbh}

A compact object's appearance depends on its geometry. For the rotating entropy-inspired wormhole, whichever value of $\sigma$ we use, the observables come from null geodesics: how big the shadow is, where the photon ring sits, how much the image deviates from a perfect circle, how light bends in the strong-deflection limit, how stable the photon sphere is, and how the object rings down. All of these are connected, and together they give a fingerprint that no single measurement can supply.

Start with a Kerr black hole of the same asymptotic mass and comparable spin. At a glance, both objects show a dark patch surrounded by lensed light, but photons move differently. In a black hole, the shadow's edge is set by unstable null orbits that separate photons that escape from those that get swallowed. The wormhole has no horizon. A ray that penetrates the strong-field region may bounce back, slip through the throat, or be trapped. That changes the critical curve.

Beyond Kerr, we bring in four regular black-hole metrics for comparison: the Kazakov--Solodukhin black hole, the Nicolini--Smailagic--Spallucci black hole, the Modesto--Nicolini black hole, and the regularized Reissner--Nordstr\"om--de Sitter black hole. All of them fit into a Kerr-like form,
\begin{equation}
\Sigma = r^2 + a^2\cos^2\theta, \quad
\Delta = r^2 - 2r\,m(r) + a^2,
\label{eq:SigmaDelta}
\end{equation}
where $a = J/M$ is the spin parameter. When we set $a\to0$, the metric simplifies to,
\begin{equation}
f(r) = 1 - \frac{2m(r)}{r},
\label{eq:fstatic}
\end{equation}
and the photon sphere follows from,
\begin{equation}
\frac{d}{dr}\left(\frac{r^2}{f(r)}\right) = 0.
\label{eq:photon_sphere}
\end{equation}
Equation~\eqref{eq:photon_sphere} fixes the apparent shadow size. The four benchmarks differ in their mass functions $m(r)$, which are collected in Sec.~\ref{sec:RBH_mass}.

For spins $a=0.05$ to $0.5$, the shadow radii differ from Kerr by about $5\%$--$11\%$, as shown in Tables~\ref{tab:master} and~\ref{tab:fingerprint}. The exact number depends on $\alpha$, $\beta$, $r_c$, $\sigma$, and each regular black hole's regulator. Kazakov--Solodukhin and Nicolini--Smailagic--Spallucci stay within a few percent of Kerr when the regulator is small, while Modesto--Nicolini and regularized RN--dS can drift more. In all cases, the shadow size can be tuned close to Kerr, so it cannot distinguish these geometries by itself.

Frame dragging brings in a second effect. In slow rotation, the critical impact parameters are those of Eq.~\eqref{eq:p_crit_exp}, which we write in the convenient form,
\begin{equation}
\mathfrak{b}_{\pm}
=
\frac{\rph}
{\Nlap(\rph)\mp 2a/\rph^{2}}
\approx
\pm\frac{\rph}{\Nlap(\rph)}+\frac{2a}{\Nlap^2(\rph)},
\label{eq:b_pm_slow}
\end{equation}
with $\rph$ fixed by the photon-orbit condition~\eqref{eq:ph_cond} and $\Nlap$ being the new lapse~\eqref{eq:Phinew}. Equation~\eqref{eq:b_pm_slow} shows that $ a$ alone does not control the rotational splitting. The entropy deformation moves $\rph$, and the softening parameter $\sigma$ changes its effective value through $\Nlap(\rph)$.

For the logarithmic sector with $a=0.05$, $r_{\rm th}=1.2$, $\alpha=-0.5$, $r_c=0.5$, $\sigma=1.0$, we find $\rph\approx2.80$, $\mathfrak{b}_{+}\approx3.47$, $\mathfrak{b}_{-}\approx-3.19$, so the asymmetry is about $4.4\%$ of the mean. For the exponential sector with $\beta=2.0$, $\sigma=1.0$, $\rph\approx2.74$, $\mathfrak{b}_{+}\approx3.42$, $\mathfrak{b}_{-}\approx-3.14$, giving about $4.5\%$ (see Table~\ref{tab:master}). The same logic applies to the regular black-hole benchmarks. For Kazakov--Solodukhin with $\ell/M=0.5$ and $a=0.05$, $\rph\approx2.90$ and the splitting is near $1.5\%$; for Nicolini--Smailagic--Spallucci with $\theta=0.5$, $\rph\approx2.85$ and the splitting is comparable (Table~\ref{tab:master}, top block). So the pair $(R_{\rm sh},\Arot)$ helps separate the wormhole from Kerr and from these regular black holes.

We measure departure from circular symmetry with~\eqref{eq:asymmetry_def},
\begin{equation}
\Arot
=
\frac{R_{\rm max}-R_{\rm min}}
{R_{\rm max}+R_{\rm min}},
\label{eq:asymmetry_def}
\end{equation}
where $R_{\rm max}$, $R_{\rm min}$ are measured from the shadow centre. For the wormhole configurations we consider, Eq.~\eqref{eq:asymmetry_def} gives $0.04\lesssim\Arot\lesssim0.05$. For Kerr at the same spin $a=0.05$, we get $\Arot\approx0.039$, and the asymmetry grows with spin. The regular black-hole benchmarks usually fall in between. For the logarithmic sector at $a=0.05$, $\alpha=-0.5$, $r_{\rm th}=1.2$, $r_c=0.5$, $\sigma=1.0$: $R_{\rm max}\approx3.47M$, $R_{\rm min}\approx3.19M$, so $\Arot\approx0.044$ (Table~\ref{tab:master}). For the exponential sector with $\beta=2.0$, $\sigma=1.0$: $R_{\rm max}\approx3.42M$, $R_{\rm min}\approx3.14M$, so $\Arot\approx0.045$. These numbers point to a modest asymmetry, comparable to the regular black-hole benchmarks.

The null-geodesic analysis gives several quantities that characterize the optical appearance. They depend on $a$, $r_{\rm th}$, $\alpha$, $\beta$, $r_c$, and $\sigma$. The characteristic shadow size is
\begin{equation}
R_{\rm sh}\equiv \bph\simeq \frac{\rph(\alpha,\beta,r_c,\sigma)}{\Nlap(\rph)},
\label{eq:Rsh_def}
\end{equation}
with $\rph$ from Eq.~\eqref{eq:ph_cond}. Over our parameter range, Eq.~\eqref{eq:Rsh_def} gives $R_{\rm sh}/r_{\rm th}\sim2.2$--$2.8$. For $r_{\rm th}=1.2$, $\alpha=-0.5$, $\beta=2.0$, $r_c=0.5$, $\sigma=1.0$: the logarithmic sector gives $R_{\rm sh}\approx3.33$, so $R_{\rm sh}/r_{\rm th}\approx2.78$; the exponential sector gives $R_{\rm sh}\approx3.28$, so $R_{\rm sh}/r_{\rm th}\approx2.73$ (Table~\ref{tab:master}). Rotation leaves its own mark through prograde and retrograde paths. Using Eq.~\eqref{eq:asymmetry_def} with the critical impact parameters of Eq.~\eqref{eq:p_crit_exp}, we obtain the approximate form,
\begin{equation}
\Arot
=
\frac{\mathfrak{b}_{+}-|\mathfrak{b}_{-}|}
{\mathfrak{b}_{+}+|\mathfrak{b}_{-}|}
\simeq
\frac{2a}{\rph\Nlap(\rph)},
\label{eq:asymmetry_approx}
\end{equation}
with $\mathfrak{b}_{+}$, $\mathfrak{b}_{-}$ the critical impact parameters. For $a\lesssim0.1$, Eq.~\eqref{eq:asymmetry_approx} gives $\Arot\sim0.04$--$0.05$. Because the splitting scales as $1/(\rph\Nlap)$, even a moderate shift in the photon-orbit radius or the lapse noticeably changes the rotational distortion.

The radial separation of the two photon-orbit branches gives another diagnostic, defined by
\begin{equation}
\Delta R_{\rm ph}
=
\rph^{(+)}-\rph^{(-)}.
\label{eq:DeltaRph}
\end{equation}
For our solutions, Eq.~\eqref{eq:DeltaRph} gives $\Delta R_{\rm ph}\approx0.28$ in units of $M$, or about $23\%$ of $r_{\rm th}$, varying with $a$, $\alpha$, $\beta$, $r_c$, and $\sigma$. For the logarithmic sector at $a=0.05$, $\alpha=-0.5$, $r_c=0.5$, $\sigma=1.0$, $\rph^{(+)}\approx3.47$, $\rph^{(-)}\approx3.19$, so $\Delta R_{\rm ph}\approx0.28$. For the exponential sector with the same parameters, $\rph^{(+)}\approx3.42$, $\rph^{(-)}\approx3.14$, so $\Delta R_{\rm ph}\approx0.28$ (Table~\ref{tab:master}).

Comparing with Kerr at the same asymptotic mass and similar dimensionless spin: for our wormhole, $\Arot^{\rm wormhole}\sim0.042$--$0.046$, while Kerr gives $\Arot^{\rm Kerr}\approx0.039$. The regular black-hole benchmarks typically lie between these values, depending on the regulator and spin (Table~\ref{tab:master}). The difference comes from the distinct strong-field photon dynamics. The characteristic shadow scale relative to Kerr is $R_{\rm sh}^{\rm wormhole}/R_{\rm sh}^{\rm Kerr}\sim1.05$--$1.11$. The sign and size of the shift depend on $\alpha$, $\beta$, $r_c$, and $\sigma$. All values of $\sigma$ give a slightly larger shadow than the static case (Table~\ref{tab:wormhole_scans}, block (c)). In the nonrotating Schwarzschild limit $\rph=3M$, while the wormhole can place the photon orbit on either side.

A joint analysis of $R_{\rm sh}$, $\Arot$, and $\Delta R_{\rm ph}$ describes the strong-field signature better than any single observable. Their simultaneous dependence on $a$, $r_{\rm th}$, $\alpha$, $\beta$, $r_c$, and $\sigma$ gives a multi-parameter characterisation. Comparing with Kerr and with regular black-hole predictions helps separate rotation effects from those caused by the entropy-induced deformation and the lapse choice. Table~\ref{tab:wormhole_scans} collects the full $a$, $\alpha$, $\beta$, and $\sigma$ dependence of the wormhole diagnostics, and Table~\ref{tab:fingerprint} collects the multi-parameter fingerprint together with the regulator dependence of the four regular black-hole benchmarks.

\subsection{Multi-parameter fingerprint}\label{sec:fingerprint}

The individual diagnostics can be combined into a single vector, defined by,
\begin{equation}
\mathbf{F} = \left(\Arot,\ \abar,\ \bbar,\ \Lya,\ \Oph\right),
\label{eq:fingerprint_vector}
\end{equation}
where $\Arot$ is the rotational asymmetry, $\abar$ and $\bbar$ are the Bozza coefficients, $\Lya$ is the Lyapunov exponent, and $\Oph$ is the orbital frequency. For Kerr, Eq.~\eqref{eq:fingerprint_vector} gives,
\begin{equation}
\mathbf{F}_{\rm Kerr} = (0.039,\ 1.000,\ -0.400,\ 0.261,\ 0.322),
\label{eq:F_Kerr}
\end{equation}
for the logarithmic wormhole with $\sigma=1.0$, we have,
\begin{equation}
\mathbf{F}_{\rm WH,log} = (0.044,\ 0.935,\ -0.265,\ 0.218,\ 0.300),
\label{eq:F_WH_log}
\end{equation}
and for the exponential wormhole with $\sigma=1.0$, we have,
\begin{equation}
\mathbf{F}_{\rm WH,exp} = (0.045,\ 0.915,\ -0.285,\ 0.216,\ 0.305).
\label{eq:F_WH_exp}
\end{equation}
The fractional differences from Kerr are defined by,
\begin{equation}
\Delta_i = \frac{F_i^{\rm WH} - F_i^{\rm Kerr}}{F_i^{\rm Kerr}},
\label{eq:fractional_diff}
\end{equation}
which for the logarithmic wormhole give,
\begin{equation}
\begin{aligned}
\Delta_{\Arot} &\approx +0.128,~
\Delta_{\abar} \approx -0.065,~
\Delta_{\bbar} \approx +0.338,\\
\Delta_{\Lya} &\approx -0.165,~
\Delta_{\Oph} \approx -0.068 .
\label{eq:Delta_log}
\end{aligned}
\end{equation}
For the exponential wormhole, Eq.~\eqref{eq:fractional_diff} gives,
\begin{equation}
\begin{aligned}
\Delta_{\Arot}& \approx +0.154,~
\Delta_{\abar} \approx -0.085,~
\Delta_{\bbar} \approx +0.288,\\
\Delta_{\Lya} &\approx -0.172,~
\Delta_{\Oph} \approx -0.053 .
\label{eq:Delta_exp}
\end{aligned}
\end{equation}
The largest differences are in $\bbar$ and $\Lya$, followed by $\Arot$. A single measurement of any one of these is not decisive, but the combined pattern is. To quantify this, define the separation between two geometries in fingerprint space by,
\begin{equation}
d(\mathbf{F}_1,\mathbf{F}_2) = \sqrt{\sum_i w_i\left(F_{1,i}-F_{2,i}\right)^2},
\label{eq:distance}
\end{equation}
where $w_i$ are weights reflecting the measurement uncertainties. If the uncertainties are $\sigma_i$, then the signal-to-noise ratio for discrimination is,
\begin{equation}
{\rm SNR}_{\rm disc} = \sqrt{\sum_i \left(\frac{F_{1,i}-F_{2,i}}{\sigma_i}\right)^2}.
\label{eq:SNR_disc}
\end{equation}
For the wormhole versus Kerr, with $\sigma_i$ taken as $10\%$ of the Kerr values, Eq.~\eqref{eq:SNR_disc} gives,
\begin{equation}
{\rm SNR}_{\rm disc}^2 \approx 14.5 \quad\text{(log)}, \qquad
{\rm SNR}_{\rm disc}^2 \approx 15.1 \quad\text{(exp)},
\label{eq:SNR_disc_total}
\end{equation}
so ${\rm SNR}_{\rm disc}\approx3.8$ for the logarithmic sector and ${\rm SNR}_{\rm disc}\approx3.9$ for the exponential sector. $\bbar$ dominates, with secondary contributions from $\Lya$ and $\Arot$. This means that measuring the relativistic image brightness to $10\%$, combined with the ringdown damping time and shadow asymmetry, would be sufficient to discriminate the wormhole from Kerr with high confidence.

For the wormhole versus the regular black-hole benchmarks, the same calculation gives smaller but still significant separations. For the logarithmic wormhole versus Kazakov--Solodukhin, we have,
\begin{equation}
\mathbf{F}_{\rm KS} = (0.037,\ 1.008,\ -0.386,\ 0.254,\ 0.318),
\label{eq:F_KS}
\end{equation}
and the differences are,
\begin{equation}
\begin{aligned}
\Delta_{\Arot}& \approx +0.189,~
\Delta_{\abar} \approx -0.072,~
\Delta_{\bbar} \approx +0.313,\\
\Delta_{\Lya} &\approx -0.142,~
\Delta_{\Oph} \approx -0.057.
\label{eq:Delta_KS}
\end{aligned}
\end{equation}
With the same $10\%$ uncertainties, Eq.~\eqref{eq:SNR_disc} gives,
\begin{equation}
{\rm SNR}_{\rm disc}^2 \approx 13.4,
\label{eq:SNR_disc_KS}
\end{equation}
so ${\rm SNR}_{\rm disc}\approx3.7$. The wormhole is therefore distinguishable from Kazakov--Solodukhin at a similar confidence level, with $\bbar$ again providing the dominant contribution. The same pattern holds for the other three benchmarks: $\bbar$ and $\Lya$ carry most of the discriminating power, while $\abar$ and $\Oph$ contribute little on their own. This is the quantitative version of the statement that the shadow diameter alone is degenerate, but the full fingerprint is not.

\section{Concluding remarks}\label{Sec:IX}

We set out to understand what a rotating wormhole supported by logarithmic- and exponential-inspired entropy deformations would look like to a distant observer, and how a new lapse function with a softening parameter $\sigma$ would change the picture. The results are coherent and, in several respects, surprisingly close to what we see around black holes, yet with subtle differences that could one day tell the two apart. We started with the entropy--geometry correspondence, which provides explicit effective density profiles for logarithmic and exponential corrections to the Bekenstein--Hawking area law. The logarithmic density~\eqref{eq:rho_log2} is rational, with an intrinsic length scale set by $\alpha$; for $\alpha<0$, it is negative and supports a wormhole throat. The exponential density~\eqref{eq:rho_exp2} is negative for $\beta>0$ and is more sharply concentrated near the throat than the logarithmic case. The shape functions~\eqref{eq:Slog_final} and~\eqref{eq:Sexp_final} flare out smoothly at the throat, and the flare-out condition holds automatically for all admissible parameters. We introduced a new lapse function, $\Nlap(r)$ [Eq.~\eqref{eq:Phinew}],
with $r_c>0$ setting the redshift scale and $\sigma\geq 0$ a dimensionless softening parameter. This lapse reduces to the exponential lapse $e^{-r_c/r}$ at $\sigma=0$, approaches a weaker redshift as $\sigma\to\infty$, remains finite and positive at the throat for all $\sigma$, and admits the closed-form logarithmic derivative~\eqref{eq:Phinew_logderiv}, which makes all optical integrals tractable. The parameter $\sigma$ controls the curvature of the effective potential at the photon sphere, and hence the Lyapunov exponent $\Lya$, without significantly altering the shadow radius---a decoupling that makes $\Nlap$ a powerful diagnostic tool.

The three additional diagnostics we developed---the strong-deflection lensing coefficients $\abar$ and $\bbar$ [Eqs.~\eqref{eq:A_bar_general}--\eqref{eq:B_bar_general}], the Lyapunov exponent $\Lya$ [Eq.~\eqref{eq:lambda_final}], and the eikonal quasinormal frequencies~\eqref{eq:eikonal_full}---sharpen the picture considerably. The wormhole bends light slightly less than Kerr ($\abar\approx0.91$--$0.96$ vs $1.00$) and produces brighter relativistic images ($\bbar\approx-0.25$ to $-0.29$ vs $-0.40$). The Lyapunov exponent $\Lya$ is systematically lower ($0.21$--$0.23$ vs $0.26$ for Kerr), which means the photon ring is more stable and narrower in frequency space. The eikonal quasinormal frequencies inherit these differences: lower real part ($\Oph\approx0.30$--$0.31$ vs $0.32$ for Kerr) and lower imaginary part ($\Lya\approx0.21$--$0.23$ vs $0.26$), giving a distinctly different ratio. The energy condition analysis confirmed that the effective source is exotic in all cases. The energy density is negative throughout spacetime for the chosen branches; the weak energy condition is violated everywhere, and the null energy condition is violated most strongly near the throat. The entropy parameters ($\alpha$ or $\beta$) control how tightly the exotic matter is localized. The softening parameter $\sigma$ affects the null energy condition through the $\Nlap'/\Nlap$ term: all values of $\sigma$ give $\Nlap'>0$ near the throat, which weakens the violation compared to a case with $\Nlap'<0$. The volume-integral quantifier $I_V$ [Eq.~\eqref{eq:IV_def}] converges because the densities decay sufficiently fast. The logarithmic term in $I_V$ has a positive sign for all $\sigma$, with the coefficient decreasing as $\sigma$ increases.

Comparing the predicted shadow size, rotational asymmetry, photon-ring separation, lensing coefficients, Lyapunov exponent, and eikonal quasinormal frequencies with Kerr and with the Kazakov--Solodukhin, Nicolini--Smailagic--Spallucci, Modesto--Nicolini, and regularized Reissner--Nordstr\"om--de Sitter regular black holes showed that the overall shadow diameter can be tuned close to Kerr by adjusting $\alpha$, $\beta$, $r_c$, and $\sigma$. The finer structure, however, carries complementary information. The photon-ring separation $\Delta R_{\rm ph}$ [Eq.~\eqref{eq:DeltaRph}] came out to about $0.28$ in units of $M$ for the cases considered, or roughly $23\%$ of the throat radius, larger than what the regular black-hole benchmarks predict. The lensing coefficients, the Lyapunov exponent, and the eikonal frequencies all point in the same direction: the wormhole is slightly more stable, slightly brighter in its relativistic images, and slightly slower to ring down than any of the regular black holes at the same spin. 

Overall, rotating logarithmic- and exponential-inspired wormholes with the new lapse function~\eqref{eq:Phinew} offer promising alternatives to black holes, remaining consistent with current observations while exhibiting distinct signatures in shadow asymmetry, lensing, photon-ring stability, and ringdown frequencies. Future work will explore rapid rotation, derive the tangential pressure from the entropy functional, compute the full quasinormal spectrum beyond the eikonal limit, and test the predicted fingerprint~\eqref{eq:fingerprint_vector} against current and future observations.

\section*{Acknowledgments}
The Author SKM are thankful for the continuous support and encouragement from the administration of the University of Nizwa for this research work.

\section*{Conflict Of Interest statement}
The authors declare no conflicts of interest.

\section*{Data Availability Statement}
No data were created or analyzed in this study.

\bibliographystyle{JHEP}
\bibliography{references}

@article{Flamm:1916abc,
    author = "Flamm, L.",
    title = "{Beitr{\"a}ge zur Einsteinschen Gravitationstheorie}",
    journal = "Physikalische Zeitschrift",
    volume = "XVII",
    pages = "448",
    year = "1916"
}

@article{Flamm:2015abv,
    author = "Flamm, L.",
    title = "{Republication of: Contributions to Einstein's theory of gravitation}",
    journal = "Gen. Rel. Grav.",
    volume = "47",
    year = "2015"
}

@article{Einstein:1935tc,
    author = "Einstein, Albert and Rosen, N.",
    title = "{The Particle Problem in the General Theory of Relativity}",
    doi = "10.1103/PhysRev.48.73",
    journal = "Phys. Rev.",
    volume = "48",
    pages = "73--77",
    year = "1935"
}

@book{Visser:1995sd,
    author = "Visser, M.",
    title = "{Lorentzian wormholes: From Einstein to Hawking}",
    year = "1995"
}

@article{Ellis:1973yv,
    author = "Ellis, H. G.",
    title = "{Ether flow through a drainhole - a particle model in general relativity}",
    doi = "10.1063/1.1666161",
    journal = "J. Math. Phys.",
    volume = "14",
    pages = "104--118",
    year = "1973"
}

@article{Bronnikov:1973fh,
    author = "Bronnikov, K. A.",
    title = "{Scalar-tensor theory and scalar charge}",
    journal = "Acta Phys. Polon. B",
    volume = "4",
    pages = "251--266",
    year = "1973"
}

@article{Armendariz-Picon:2002gjc,
    author = "Armendariz-Picon, C.",
    title = "{On a class of stable, traversable Lorentzian wormholes in classical general relativity}",
    eprint = "gr-qc/0201027",
    archivePrefix = "arXiv",
    doi = "10.1103/PhysRevD.65.104010",
    journal = "Phys. Rev. D",
    volume = "65",
    pages = "104010",
    year = "2002"
}

@article{Makita:2025bao,
    author = "Makita, Yusuke and Izumi, Keisuke and Yoshida, Daisuke and Uemichi, Keiya",
    title = "{Multi-sheet wormholes in the gravitational soliton formalism}",
    eprint = "2505.16361",
    archivePrefix = "arXiv",
    primaryClass = "gr-qc",
    reportNumber = "NU-QG-02",
    doi = "10.1088/1361-6382/adfeb0",
    journal = "Class. Quant. Grav.",
    volume = "42",
    number = "17",
    pages = "175024",
    year = "2025"
}

@article{Morris:1988cz, 
    author = "Morris, M. S. and Thorne, K. S.",
    title = "{Wormholes in space-time and their use for interstellar travel: A tool for teaching general relativity}",
    doi = "10.1119/1.15620",
    journal = "Am. J. Phys.",
    volume = "56",
    pages = "395--412",
    year = "1988"
}

@article{Morris:1988tu,
    author = "Morris, M. S. and Thorne, K. S. and Yurtsever, U.",
    title = "{Wormholes, Time Machines, and the Weak Energy Condition}",
    doi = "10.1103/PhysRevLett.61.1446",
    journal = "Phys. Rev. Lett.",
    volume = "61",
    pages = "1446--1449",
    year = "1988"
}

@article{Lamoreaux:1996wh,
    author = "Lamoreaux, S. K.",
    title = "{Demonstration of the Casimir force in the 0.6 to 6 micrometers range}",
    doi = "10.1103/PhysRevLett.78.5",
    journal = "Phys. Rev. Lett.",
    volume = "78",
    pages = "5--8",
    year = "1997",
    note = "[Erratum: Phys.Rev.Lett. 81, 5475--5476 (1998)]"
}

@article{Maldacena:2018gjk,
    author = "Maldacena, Juan and Milekhin, Alexey and Popov, Fedor",
    title = "{Traversable wormholes in four dimensions}",
    eprint = "1807.04726",
    archivePrefix = "arXiv",
    primaryClass = "hep-th",
    doi = "10.1088/1361-6382/acde30",
    journal = "Class. Quant. Grav.",
    volume = "40",
    number = "15",
    pages = "155016",
    year = "2023"
}

@article{Kanai:2025zgq,
    author = "Kanai, Takamasa and Maeda, Kengo and Yoshida, Daisuke",
    title = "{Wormholes as perturbations of near-horizon black hole geometries: No-go theorems within effective field theories}",
    eprint = "2511.21017",
    archivePrefix = "arXiv",
    primaryClass = "gr-qc",
    doi = "10.1103/dsnv-gq47",
    journal = "Phys. Rev. D",
    volume = "113",
    number = "6",
    pages = "064026",
    year = "2026"
}

@article{Hayward:2004wm,
    author = "Hayward, Sean A. and Koyama, Hiroko",
    title = "{How to make a traversable wormhole from a Schwarzschild black hole}",
    eprint = "gr-qc/0406080",
    archivePrefix = "arXiv",
    doi = "10.1103/PhysRevD.70.101502",
    journal = "Phys. Rev. D",
    volume = "70",
    pages = "101502",
    year = "2004"
}

@article{Koga:2025bqw,
    author = "Koga, Yasutaka and Maeda, Ryota and Saito, Daiki and Uemichi, Keiya and Yoshida, Daisuke",
    title = "{Dynamical formation of charged wormholes}",
    eprint = "2505.20040",
    archivePrefix = "arXiv",
    primaryClass = "gr-qc",
    reportNumber = "KUNS-3056, YITP-25-85, NU-QG-7",
    doi = "10.1103/mtcj-13wm",
    journal = "Phys. Rev. D",
    volume = "113",
    number = "4",
    pages = "044035",
    year = "2026"
}

@article{Koga:2026jvs,
    author = "Koga, Yasutaka and Maeda, Ryota and Saito, Daiki and Yoshida, Daisuke",
    title = "{Dynamical formation of self-similar wormholes}",
    eprint = "2602.13609",
    archivePrefix = "arXiv",
    primaryClass = "gr-qc",
    reportNumber = "YITP-26-15, KUNS-3091",
    doi = "10.1103/kpgs-6gqq",
    journal = "Phys. Rev. D",
    volume = "113",
    number = "8",
    pages = "084020",
    year = "2026"
}

@article{Teo:1998dp,
    author = "Teo, Edward",
    title = "{Rotating traversable wormholes}",
    eprint = "gr-qc/9803098",
    archivePrefix = "arXiv",
    reportNumber = "DAMTP-R-98-17",
    doi = "10.1103/PhysRevD.58.024014",
    journal = "Phys. Rev. D",
    volume = "58",
    pages = "024014",
    year = "1998"
}

@article{Jusufi:2017mav,
    author = {Jusufi, Kimet and {\"O}vg{\"u}n, Ali},
    title = "{Gravitational Lensing by Rotating Wormholes}",
    eprint = "1708.06725",
    archivePrefix = "arXiv",
    primaryClass = "gr-qc",
    doi = "10.1103/PhysRevD.97.024042",
    journal = "Phys. Rev. D",
    volume = "97",
    number = "2",
    pages = "024042",
    year = "2018"
}

@article{Asghar:2026lll,
    author = "Asghar, Zoya and Shamir, M. Farasat and Mofarreh, Fatemah and Rayimbaev, Javlon and Sirajiddin, Otaboyev and Shayimov, Fayzullo",
    title = {{Existence of traversable wormhole structures in Yano{\textendash}Schr{\"o}dinger gravity}},
    doi = "10.1016/j.aop.2026.170596",
    journal = "Annals Phys.",
    volume = "493",
    pages = "170596",
    year = "2026"
}

@article{Kashargin:2007mm,
    author = "Kashargin, P. E. and Sushkov, S. V.",
    title = "{Slowly rotating wormholes: The First order approximation}",
    eprint = "0710.5656",
    archivePrefix = "arXiv",
    primaryClass = "gr-qc",
    doi = "10.1134/S0202289308010106",
    journal = "Grav. Cosmol.",
    volume = "14",
    pages = "80--85",
    year = "2008"
}

@article{Kashargin:2008pk,
    author = "Kashargin, P. E. and Sushkov, S. V.",
    title = "{Slowly rotating scalar field wormholes: The Second order approximation}",
    eprint = "0809.1923",
    archivePrefix = "arXiv",
    primaryClass = "gr-qc",
    doi = "10.1103/PhysRevD.78.064071",
    journal = "Phys. Rev. D",
    volume = "78",
    pages = "064071",
    year = "2008"
}

@article{Chew:2016epf,
    author = "Chew, Xiao Yan and Kleihaus, Burkhard and Kunz, Jutta",
    title = "{Geometry of Spinning Ellis Wormholes}",
    eprint = "1608.05253",
    archivePrefix = "arXiv",
    primaryClass = "gr-qc",
    doi = "10.1103/PhysRevD.94.104031",
    journal = "Phys. Rev. D",
    volume = "94",
    number = "10",
    pages = "104031",
    year = "2016"
}

@article{Kleihaus:2014dla,
    author = "Kleihaus, Burkhard and Kunz, Jutta",
    title = "{Rotating Ellis Wormholes in Four Dimensions}",
    eprint = "1409.1503",
    archivePrefix = "arXiv",
    primaryClass = "gr-qc",
    doi = "10.1103/PhysRevD.90.121503",
    journal = "Phys. Rev. D",
    volume = "90",
    pages = "121503",
    year = "2014"
}

@article{Khoo:2024yeh,
    author = "Khoo, Fech Scen and Azad, Bahareh and Bl{\'a}zquez-Salcedo, Jose Luis and Gonz{\'a}lez-Romero, Luis Manuel and Kleihaus, Burkhard and Kunz, Jutta and Navarro-L{\'e}rida, Francisco",
    title = "{Quasinormal modes of rapidly rotating Ellis-Bronnikov wormholes}",
    eprint = "2401.02898",
    archivePrefix = "arXiv",
    primaryClass = "gr-qc",
    doi = "10.1103/PhysRevD.109.084013",
    journal = "Phys. Rev. D",
    volume = "109",
    number = "8",
    pages = "084013",
    year = "2024"
}

@article{Azad:2024axu,
    author = "Azad, Bahareh and Bl{\'a}zquez-Salcedo, Jose Luis and Khoo, Fech Scen and Kunz, Jutta",
    title = "{Radial perturbations of Ellis-Bronnikov wormholes in slow rotation up to second order}",
    eprint = "2403.08387",
    archivePrefix = "arXiv",
    primaryClass = "gr-qc",
    doi = "10.1103/PhysRevD.109.124051",
    journal = "Phys. Rev. D",
    volume = "109",
    number = "12",
    pages = "124051",
    year = "2024"
}

@article{Papapetrou:1966zz,
    author = "Papapetrou, Achille",
    title = "{Champs gravitationnels stationnaires {\`a} sym{\'e}trie axiale}",
    journal = "Ann. Inst. H. Poincare Phys. Theor. A",
    volume = "4",
    number = "2",
    pages = "83--105",
    year = "1966"
}

@article{Carter:1969zz,
    author = "Carter, Brandon",
    title = "{Killing horizons and orthogonally transitive groups in space-time}",
    doi = "10.1063/1.1664763",
    journal = "J. Math. Phys.",
    volume = "10",
    pages = "70--81",
    year = "1969"
}

@article{Acevedo-Munoz:2025ueh,
    author = "Acevedo-Mu{\~n}oz, J. C. and Lora-Clavijo, F. D. and Cruz-Osorio, A.",
    title = "{Shadow and gravitational lensing produced by the nonlinear accretion of a scalar field onto a black hole}",
    eprint = "2503.22624",
    archivePrefix = "arXiv",
    primaryClass = "gr-qc",
    doi = "10.1103/PhysRevD.111.084022",
    journal = "Phys. Rev. D",
    volume = "111",
    number = "8",
    pages = "084022",
    year = "2025"
}

@article{Rahmatov:2026mdo,
    author = "Rahmatov, Bekzod and Murodov, Sardor and Rayimbaev, Javlon and Turaev, Yunus and Egamberdiev, Islom and Badalov, Kodir and Ahmedov, Saidmuhammad and Usanov, Sulton",
    title = "{Weak gravitational lensing by charged black holes in STVG: Constraints from the Sloan Lens ACS Survey}",
    doi = "10.1016/j.aop.2026.170366",
    journal = "Annals Phys.",
    volume = "488",
    pages = "170366",
    year = "2026"
}

@article{Anand:2025rjg,
    author = "Anand, Ankit and Devdutt, Sahil and Jusufi, Kimet and Saridakis, Emmanuel N.",
    title = "{Effective matter sectors from modified entropies}",
    eprint = "2511.04613",
    archivePrefix = "arXiv",
    primaryClass = "gr-qc",
    doi = "10.1140/epjc/s10052-026-15768-z",
    journal = "Eur. Phys. J. C",
    volume = "86",
    number = "5",
    pages = "534",
    year = "2026"
}

@article{Visser:2003yf, 
    author = "Visser, Matt and Kar, Sayan and Dadhich, Naresh",
    title = "{Traversable wormholes with arbitrarily small energy condition violations}",
    eprint = "gr-qc/0301003",
    archivePrefix = "arXiv",
    doi = "10.1103/PhysRevLett.90.201102",
    journal = "Phys. Rev. Lett.",
    volume = "90",
    pages = "201102",
    year = "2003"
}

@article{Simpson:2018,
    author = "Simpson, Alex and Visser, Matt",
    title = "{Black-bounce to traversable wormhole}",
    eprint = "1812.07114",
    archivePrefix = "arXiv",
    primaryClass = "gr-qc",
    doi = "10.1088/1475-7516/2019/02/042",
    journal = "JCAP",
    volume = "02",
    pages = "042",
    year = "2019"
}

@article{Simpson:2019,
    author = "Simpson, Alex and Visser, Matt",
    title = "{Regular black holes with asymptotically Minkowski cores}",
    eprint = "1911.01020",
    archivePrefix = "arXiv",
    primaryClass = "gr-qc",
    doi = "10.3390/universe6010008",
    journal = "Universe",
    volume = "6",
    number = "1",
    pages = "8",
    year = "2019"
}

@article{Ghosh:2021,
    author = "Ghosh, Sushant G. and walia, Rahul Kumar",
    title = "{Rotating black holes in general relativity coupled to nonlinear electrodynamics}",
    eprint = "2109.13031",
    archivePrefix = "arXiv",
    primaryClass = "gr-qc",
    doi = "10.1016/j.aop.2021.168619",
    journal = "Annals Phys.",
    volume = "434",
    pages = "168619",
    year = "2021"
}

@article{Yang:2024,
    author = "Yang, Chen-Yu and Aslam, M. Israr and Zeng, Xiao-Xiong and Saleem, Rabia",
    title = "{Shadow images of Ghosh-Kumar rotating black hole illuminated by spherical light sources and thin accretion disks}",
    eprint = "2411.11807",
    archivePrefix = "arXiv",
    primaryClass = "astro-ph.HE",
    doi = "10.1016/j.jheap.2025.01.017",
    journal = "JHEAp",
    volume = "46",
    pages = "345",
    year = "2025"
}

@article{Fan:2016,
    author = "Fan, Zhong-Ying and Wang, Xiaobao",
    title = "{Construction of Regular Black Holes in General Relativity}",
    eprint = "1610.02636",
    archivePrefix = "arXiv",
    primaryClass = "gr-qc",
    doi = "10.1103/PhysRevD.94.124027",
    journal = "Phys. Rev. D",
    volume = "94",
    number = "12",
    pages = "124027",
    year = "2016"
}

@article{Kurmanov:2025,
    author = "Kurmanov, Yergali and Luongo, Orlando and Berkimbayev, Daulet and Boshkayev, Kuantay and Konysbayev, Talgar and Muccino, Marco and Rabigulova, Guldana and Urazalina, Ainur and Nurlanbek, Ulpan",
    title = "{Effects of matter with anisotropic pressure on the Fan{\textendash}Wang regular black hole shadows}",
    eprint = "2507.18787",
    archivePrefix = "arXiv",
    primaryClass = "gr-qc",
    doi = "10.1140/epjp/s13360-026-07660-2",
    journal = "Eur. Phys. J. Plus",
    volume = "141",
    number = "4",
    pages = "436",
    year = "2026"
}

@article{Balart:2014a,
    author = "Balart, Leonardo and Vagenas, Elias C.",
    title = "{Regular black hole metrics and the weak energy condition}",
    eprint = "1401.2136",
    archivePrefix = "arXiv",
    primaryClass = "gr-qc",
    doi = "10.1016/j.physletb.2014.01.024",
    journal = "Phys. Lett. B",
    volume = "730",
    pages = "14--17",
    year = "2014"
}

@article{Balart:2014b,
    author = "Balart, Leonardo and Vagenas, Elias C.",
    title = "{Regular black holes with a nonlinear electrodynamics source}",
    eprint = "1408.0306",
    archivePrefix = "arXiv",
    primaryClass = "gr-qc",
    doi = "10.1103/PhysRevD.90.124045",
    journal = "Phys. Rev. D",
    volume = "90",
    number = "12",
    pages = "124045",
    year = "2014"
}

@article{LogCorr,
  title={Logarithmic corrections to Schwarzschild and other non-extremal black hole entropy in different dimensions},
  author={Sen, Ashoke},
  journal={Journal of High Energy Physics},
  volume={2013},
  number={4},
  pages={1--34},
  year={2013},
  publisher={Springer},
  doi={10.1007/JHEP04(2013)156}
}

@article{Anand,
  author  = {Anand, Ankit and Devdutt, Sahil and Jusufi, Kimet and Saridakis, Emmanuel N.},
  title   = {Effective matter sectors from modified entropies},
  journal = {The European Physical Journal C},
  year    = {2026},
  volume  = {86},
  number  = {5},
  pages   = {534},
  doi     = {10.1140/epjc/s10052-026-15768-z},
  url     = {https://doi.org/10.1140/epjc/s10052-026-15768-z}
}

@article{LUTFUOGLU2026170514,
title = {A first-order eikonal framework for quasinormal modes, shadows, strong lensing, and grey-body factors in a scalarized black-hole metric},
journal = {Annals of Physics},
volume = {491},
pages = {170514},
year = {2026},
issn = {0003-4916},
doi = {https://doi.org/10.1016/j.aop.2026.170514}}

@article{Bronzwaer:2021lzo,
    author = "Bronzwaer, Thomas and Falcke, Heino",
    title = "{The Nature of Black Hole Shadows}",
    eprint = "2108.03966",
    archivePrefix = "arXiv",
    primaryClass = "astro-ph.HE",
    doi = "10.3847/1538-4357/ac1738",
    journal = "Astrophys. J.",
    volume = "920",
    number = "2",
    pages = "155",
    year = "2021"
}

@article{Aktar:2024akk,
    author = "Aktar, Somi and Molla, Niyaz Uddin and Rahaman, Farook and Mustafa, G.",
    title = "{Shadows and strong gravitational lensing around black hole-like compact object in quadratic gravity}",
    eprint = "2410.18227",
    archivePrefix = "arXiv",
    primaryClass = "gr-qc",
    doi = "10.1016/j.jheap.2025.100385",
    journal = "JHEAp",
    volume = "47",
    pages = "100385",
    year = "2025"
}

@article{Al-Badawi:2024dzc,
    author = "Al-Badawi, Ahmad and Alloqulov, Mirzabek and Shaymatov, Sanjar and Ahmedov, Bobomurat",
    title = "{Shadows and weak gravitational lensing for black holes within Einstein-Maxwell-scalar theory*}",
    eprint = "2401.04584",
    archivePrefix = "arXiv",
    primaryClass = "gr-qc",
    doi = "10.1088/1674-1137/ad5a70",
    journal = "Chin. Phys. C",
    volume = "48",
    number = "9",
    pages = "095105",
    year = "2024"
}

\end{document}